\documentclass[%
aip,
amsmath,amssymb,
preprint,%
]{revtex4-1}

\usepackage{graphicx}
\usepackage{dcolumn}
\usepackage{bm}

\usepackage[utf8]{inputenc}
\usepackage[T1]{fontenc}
\usepackage{mathptmx}
\usepackage{etoolbox}
\usepackage{color}
\usepackage{xcolor}

\makeatletter
\def\@email#1#2{%
	\endgroup
	\patchcmd{\titleblock@produce}
	{\frontmatter@RRAPformat}
	{\frontmatter@RRAPformat{\produce@RRAP{*#1\href{mailto:#2}{#2}}}\frontmatter@RRAPformat}
	{}{}
}%
\makeatother
\begin{document}
	
	\preprint{}
	
	\title[]{Dominant balance-based adaptive mesh refinement for incompressible fluid flows}
	\author{Gaurav Kumar} 
	\affiliation{University of Nevada, Reno, Nevada 89557, USA}
	
	\author{Aditya Nair}
	\affiliation{University of Nevada, Reno, Nevada 89557, USA}
	\email{adityan@unr.edu}

		\date{\today}
		
		\begin{abstract}
This work introduces a novel adaptive mesh refinement (AMR) method that utilizes dominant balance analysis (DBA) for efficient and accurate grid adaptation in computational fluid dynamics (CFD) simulations. The proposed method leverages a Gaussian mixture model (GMM) to classify grid cells into active and passive regions based on the dominant physical interactions in the equation space. Unlike traditional AMR strategies, this approach does not rely on heuristic-based sensors or user-defined parameters, providing a fully automated and problem-independent framework for AMR. Applied to the incompressible Navier-Stokes equations for unsteady flow past a cylinder, the DBA-based AMR method achieves comparable accuracy to high-resolution grids while reducing computational costs by up to 70\%. The validation highlights the method’s effectiveness in capturing complex flow features while minimizing grid cells, directing computational resources toward regions with the most critical dynamics. This modular and scalable strategy is adaptable to a range of applications, presenting a promising tool for efficient high-fidelity simulations in CFD and other multiphysics domains.

		\end{abstract}
		
		\maketitle
		
		\section{Introduction}\label{sec:Intro}
		
		In past few decades, advances in supercomputers have significantly enhanced numerical simulations for multiphysics applications, from designing automobiles and aircrafts \cite{buljac2016automobile, rizzi2021historical} to studying star and galaxy formation \cite{crain2023hydrodynamical}, understanding bird flight and fish swimming \cite{lin2016three, macias2020three}, and developing noise-free wind turbines \cite{tadamasa2011numerical}. Despite these advancements, the growing complexity of problems consistently outpaces improvements in computational speed. This gap highlights the critical need and presents a significant opportunity to develop more efficient computational tools for solving partial differential equations (PDEs).
		
		Numerically solving a PDE involves spatial and temporal discretization of the entire computational domain. The computational complexity of the problem generally reflects the range of spatio-temporal scales involved in the physical processes. For example, the computational complexity of simulating turbulent boundary layer by solving Navier--Stokes equation increases rapidly with the Reynolds number ($Re$). The number of grid points needed to simulate a developing boundary layer over a flat plate grows as $Re^{2.05}$ for direct numerical simulation (DNS) and $Re^{1.86}$ for large eddy simulation (LES) \cite{yang2021grid}. Therefore, managing computational resources effectively is crucial for accurately simulating large-scale computational problems within limited resources \cite{lucas2014doe}. 
		
		Adaptive mesh refinement (AMR) is a computational technique used to optimize resource allocation by dynamically adjusting the resolution of the computational grid according to the evolving solution \cite{berger1984adaptive}. This approach is widely utilized in computational fluid dynamics (CFD) to enhance simulation efficiency and significantly reduce computational costs. AMR techniques focus computational efforts on regions with the highest solution errors, thereby improving overall accuracy with less computational expense. However, an a-priori estimation of the solution error in a numerical simulation is a challenging task. Employing an efficient and accurate AMR technique in any CFD code requires two key components: (i) a sensor indicative of the local truncation error for the discretized differential equation at each grid cell, and (ii) determining a cluster of grid cells for further refinement based on the sensor values.
		
		The accurate estimation of truncation error is crucial to ensure the accuracy of simulations, as insufficient resolution in important parts of the computational domain can lead to inaccurate results \cite{freitas2002issue}. However, a-priori estimate of truncation error are seldom available in a simulation. Hence, several strategies are commonly used to assess the sensitivity of the overall solution error from individual grid cells as proxy for the former problem. These include error minimization techniques \cite{berger1984adaptive}, output-based error estimation \cite{muller2001solution,nemec2008adjoint,li2011continuous,hartmann2011adjoint,fidkowski2011review}, and sensor-based methods \cite{hartmann2011level}. Error minimization techniques, such as Richardson extrapolation \cite{aftosmis2002multilevel}, involve running an additional simulation on a refined grid to estimate the error based on the difference in solutions obtained at two different grid levels. This approach, while effective, incurs additional computational costs due to the need for another simulation. Output-based error estimation typically employs the adjoint method to calculate the sensitivity of each grid cell with respect to a chosen output or metric. These a posteriori computed metrics, such as the energy norm \cite{wu1990error}, heuristics \cite{leicht2010error}, or Hessian metrics \cite{pain2001tetrahedral} serve as surrogates to represent the actual errors in the simulation. However, developing an adjoint solver for error estimation is a non-trivial task, especially for unsteady simulations. Sensor-based techniques, on the other hand, utilize the gradients of solution fields, such as velocity or density, or the gradient of an interface tracking variable (as used in level-set methods) to estimate errors \cite{compere2008transient,wissink2010coupled}. These methods are more straightforward to implement compared to adjoint-based techniques but may not always capture the full complexity of the error distribution. Overall, the choice of error estimation strategy depends on the specific requirements of the simulation and the acceptable computational overhead for AMR.
		
\begin{figure*}[ht]
			\centering
			\includegraphics[width=\linewidth]{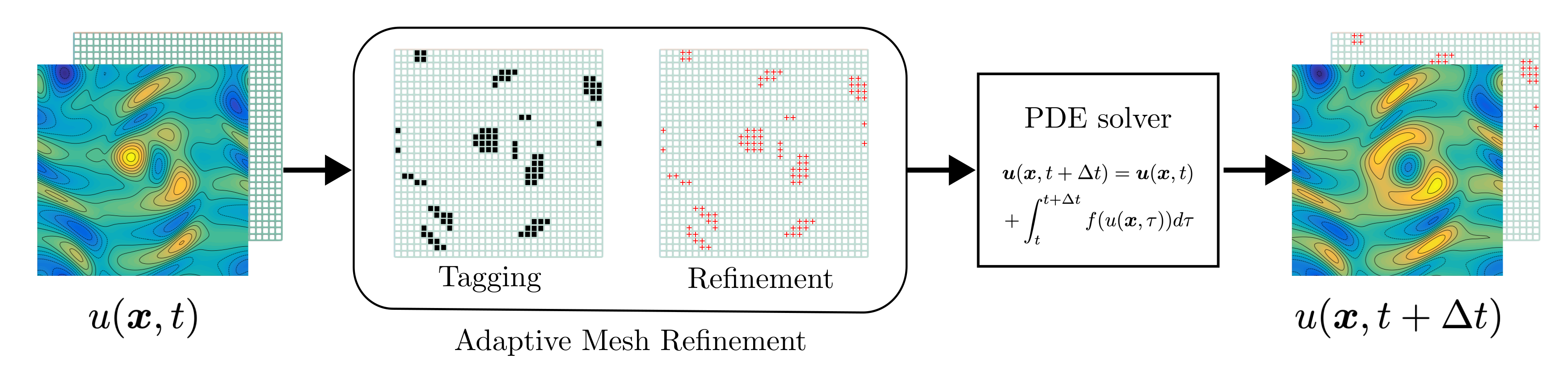}
			\caption{A schematic representation of the algorithm to solve a PDE with AMR. The input-output like involvement of AMR step in the solution algorithm emphasizes the modular nature of the AMR process which takes solution $u(\bm{x},t)$ and a grid as input and returns a new grid with appropriate refinement as output and can be embedded into any PDE solution algorithm in general.}
			\label{fig:AMR}
		\end{figure*}

		The second step in the AMR process involves identifying a cluster of grid cells for further refinement, with the objective of balancing improved accuracy against computational cost. This step can be viewed as an optimization problem, where the goal is to maximize the value of a submodular set function—such as the aggregate truncation error or sensor value across the computational domain—while selecting the fewest grid cells possible \cite{vidal2012combinatorial}. However, solving this optimization problem with numerical simulations is computationally expensive. As a result, most researchers opt for a thresholding method. This approach tags grid cells with sensor values that fall outside a predefined threshold or cutoff range. However, these threshold values are often arbitrary, relying on the user's experience with the specific simulation, and may lead to suboptimal grid refinement. Figure \ref{fig:AMR} illustrates the AMR process in a general CFD simulation. 
		
		To develop an automated Adaptive Mesh Refinement (AMR) strategy that is independent of specific problem statements and requires no user input, we adopt an interaction-based perspective \cite{taira2022network}. The foundation of this approach lies in understanding that the temporal evolution of complex systems, such as turbulent flow dynamics, can be effectively modeled through interactions among vortices, which can be represented as a graph or network, as demonstrated by Taira et al. \cite{taira2016network, meena2021identifying}. This network-based representation facilitates various processes, such as graph sparsification \cite{nair2015network}, community detection \cite{gopalakrishnan2018network, nair2023network}, and the selection of optimal features — in this context, grid cells with optimal features — leading to compressive sampling \cite{bai2019randomized} and sparse reconstruction of the actual dynamics \cite{manohar2018data}.
		
		However, constructing and analyzing a network across a computational domain with a large number of grid cells becomes computationally expensive and impractical as the grid cell count increases. To address this challenge, we employ a more efficient interaction-based analysis in the equation space, inspired by the dominant balance analysis (DBA) method presented by Callaham et al. \cite{callaham2021learning}. Callaham et al. introduced a data-driven framework for identifying dominant balance regimes within complex systems using an unsupervised learning approach. This method involves creating an equation space, where different terms in the governing equations are evaluated and clustered to identify regions dominated by specific terms \cite{callaham2021learning}.
		
		By applying this interaction-based analysis in the equation space, we can identify regions with dominant physical processes, which in turn guide the refinement process more effectively. This approach allows for a more efficient and targeted AMR strategy by focusing computational resources on regions where interactions are most significant. Moreover, these high-interaction regions of the flow, which have a substantial influence on the overall dynamics, become the focal points of our refinement strategy, ensuring that computational efforts are concentrated where they are most needed. This interaction-based perspective, therefore, ties together the entire AMR process, enabling an automated and problem-independent refinement strategy that enhances both accuracy and efficiency in complex simulations.
		
		This work introduces a generalizable grid tagging criterion based on identifying regions with active physical processes within the computational domain. Integrating dominant balance into AMR offers several benefits: it allows the refinement algorithm to adapt dynamically to the evolving solution, requires no user input or prior knowledge about the solution, and ensures that computational resources are allocated to the regions where they are most needed. Additionally, it provides a robust and general tagging strategy that can be applied to solving any partial differential equation. The detailed description of the dominant balance analysis (DBA) method and its application in the context of AMR is provided in section \ref{sec:Method}. In section \ref{sec:validation}, we demonstrate the application of DBA in automatically tagging the important regions of the flow domain through steady and unsteady two-dimensional incompressible flow over a cylinder. In section \ref{sec:Results and analysis}, we apply DBA for 3D incompressible turbulent flow over a cylinder at $Re = 3900$ to demonstrate its efficacy in adaptive mesh refinement. Concluding remarks are offered in section \ref{sec:Conclusion}.
		
		\section{Method}\label{sec:Method}
		
		In this section, we describe the proposed method in the context of solving the highly non-linear incompressible Navier-Stokes equation, a well-known computationally challenging PDE in the research community. In the present article, we focus on the application of the method in solving incompressible fluid flows. However, due to the generic nature of the proposed method and the modular implementation of the algorithm, it can be naturally extended for any PDE solution.
		
		The Navier-Stokes equation for an incompressible fluid flow is given as follows:
		\begin{equation}
			\bm{R}(\bm{x},t) ~\equiv~ \underbrace{(\bm{u}\cdot\nabla)\bm{u}}_{Q_1} ~+~ \underbrace{\nabla \left(\frac{p}{\rho}\right)}_{Q_2} ~-~ \underbrace{\frac{1}{Re}\nabla^2\bm{u}}_{Q_3} ~+~ \underbrace{\frac{\partial \vec{u}}{\partial t}}_{Q_4} ~=~ 0,
			\label{eq:governingEqn}
		\end{equation}
		where $\bm{R}$ is the residual of the governing equation and $\bm{u}, ~p$ are velocity and pressure fields, repectively over spatial-coordinates ($\bm{x}$) and time-coordinate ($t$). In this equation, Reynolds number $Re = (UD/\nu)$, with free-stream flow speed $U$, characteristic length $D$ and kinematic viscosity $\nu$, is the non-dimensional parameter leading to different flow physics. The solution field also satisfies mass conservation leading to an additional constraint equation of the form $\nabla\cdot\bm{u} ~=~ 0$.
		
		In a numerical algorithm for solving equation \ref{eq:governingEqn}, the goal is to compute the discrete solution fields $\bm{u}(\bm{x},t)$ and $p(\bm{x},t)$ within the computational domain such that the residual $\bm{R}(\bm{x},t)$ equals zero at each grid cell, within machine-precision tolerance. This essentially means that at every grid cell, a balance occurs between different processes represented by $Q_i$s such that $\sum Q_i = 0$. Based on this, we can infer that if some or most of $Q_i$s are negligibly small at a grid cell, the cell is likely to have a minimal impact on the overall dynamics of the solution. In contrast, a grid point with a greater number of significant $Q_i$s balancing each other is more likely to influence the dynamics. We call the former kind of grid cells passive and latter kind of grid cells active. When employing an AMR strategy in a numerical computation, objective is to tag and efficiently cluster together all the active grid cells for successive grid refinement. This process is similar to identifying important vortical features in the computational domain that can sparsely represent the dynamics of the actual system \cite{taira2022network}; albeit vortical features might not be the only important feature to track for different flow phenomena \cite{menon2021significance} in a general PDE.

		DBA \cite{callaham2021learning} provides an inexpensive method to identify the active grid cells in a computational domain. In DBA, we consider the equation space formed by the terms of the governing equation: $Q_{i = 1,2,3,4}$ in equation \ref{eq:governingEqn}. Each term is evaluated at every grid cell representing a contribution in locally balancing the governing equation. For example, in equation \ref{eq:governingEqn}, the terms $Q_1$, $Q_2$, $Q_3$ and $Q_4$ represent contribution of local convection, pressure gradient, diffusion and inertia to the overall balance of linear momentum in the flow, respectively. This can be understood as a network of interacting terms in the equation space at each grid cell. A dominant balance regime is defined as a region where a subset of the terms in the governing equation approximately balance each other \cite{callaham2021learning}.

		 The natural geometric interpretation in equation space allows DBA to employ standard machine learning tools to automatically identify regions with high levels of interaction/activity \cite{callaham2021learning}. To identify grid cells for AMR, the dominant balance (active) region in the computational domain can be identified as the cluster of grid cells with significant covariance in several directions of the equation space. We choose to cluster grid points into active and passive regions using Gaussian mixture models (GMM)\cite{callaham2021learning}. GMM is computationally less expensive to train and assumes that the probability density function of each data point ( $Q_i$ ) is a weighted sum of K Gaussian distributions, expressed as:
		 \begin{equation}
			 p(x_i) = \sum_{k=1}^{K} \pi_k \mathcal{N}(Q_i \mid \bm{\mu}_k, \bm{\Sigma}_k),
			 \end{equation}
		 where $\pi_k$ denotes the mixing coefficient for the $k$-th Gaussian component, satisfying $\sum_{k=1}^{K} \pi_k = 1$ and $\pi_k \geq 0$. The term $\mathcal{N}(Q_i \mid \bm{\mu}_k, \bm{\Sigma}_k)$ represents a Gaussian distribution with mean $\bm{\mu}_k$ and covariance matrix $\bm{\Sigma}_k$, given by:
		 \begin{equation}
			 \mathcal{N}(Q_i \mid \bm{\mu}_k, \bm{\Sigma}_k) = \frac{1}{(2\pi)^{d/2} |\bm{\Sigma}_k|^{1/2}} \exp\left[ -\frac{1}{2} (Q_i - \bm{\mu}_k)^\top \bm{\Sigma}_k^{-1} (Q_i - \bm{\mu}_k)\right].
			 \end{equation}
		 where $d$ is the dimension of the input data points. In the present work $d = 4$.
		
		 The off-diagonal elements of the covariance matrix $\bm{\Sigma}_k$ capture the degree of relative variability between different terms in the equation. Higher values in these off-diagonal elements suggest stronger interactions between the corresponding terms, implying that the behavior of one term is significantly influenced by others. These interactions often highlight regions in the domain where the flow dynamics are complex and potentially under-resolved. Clusters with covariance matrices exhibiting large number of off-diagonal elements correspond to regions where the governing dynamics are highly coupled and, therefore, more sensitive to numerical resolution. Refining the mesh in these regions ensures that the numerical method can accurately capture these interactions, thereby enhancing the overall fidelity of the simulation.

		The above described DBA based AMR strategy is implemented through python bindings of an incompressible flow solver in OpenFOAM CFD toolbox \cite{ofv2312} via a lightweight header-only library \emph{pybind11} \cite{pybind11}. This allows for flexible algorithm to implement and test clustering algorithms in python while seamlessly communicating with pre-existing numerical solution algorithms and grid refinement libraries in OpenFOAM. 
		
		\section{Validation}\label{sec:validation}
		
		The performance of the proposed AMR strategy hinges on two key factors: (i) the efficient tagging of grid cells for refinement, and (ii) the dynamic adaptation of the grid tagging strategy across varying levels of refinement and evolving solutions over time. AMR is particularly valuable in unsteady flow simulations, where a static grid cannot maintain optimal resolution as the solution field changes. A practical approach is to perform dominant balance analysis (DBA) on a coarse grid and reuse the pre-trained Gaussian Mixture Model (GMM) on an adaptively refined grid to capture the unsteady solution, thereby reducing the computational cost of training a new GMM at each time step. 
		
		In this section, we demonstrate the effectiveness of DBA in meeting these requirements by illustrating its ability to accurately tag high-interaction regions in an interpretable manner. We begin with a two-dimensional, steady, incompressible flow past a circular cylinder at $Re = 40$. To examine the GMM’s applicability across different grid refinements and time-dependent solution fields, we apply the same setup at $ Re = 100$, where the flow exhibits unsteady behavior. Specifically, we assess (i) the impact of flow unsteadiness on DBA’s ability to identify active grid cells, and (ii) the effectiveness of using a pre-trained GMM from the coarse grid on a refined grid.

		All simulations in this work are performed using an incompressible flow solver called PimpleFoam in OpenFOAM CFD toolbox \cite{ofv2312} which is a finite volume solver equipped with several libraries for turbulence modeling and AMR. The numerical discretization is overall $2^{nd}$ order accurate in space and time integration is performed using a $1^{st}$ order implicit Euler scheme. 
		
\begin{figure}[ht]
			\centering
			\includegraphics[width=\linewidth]{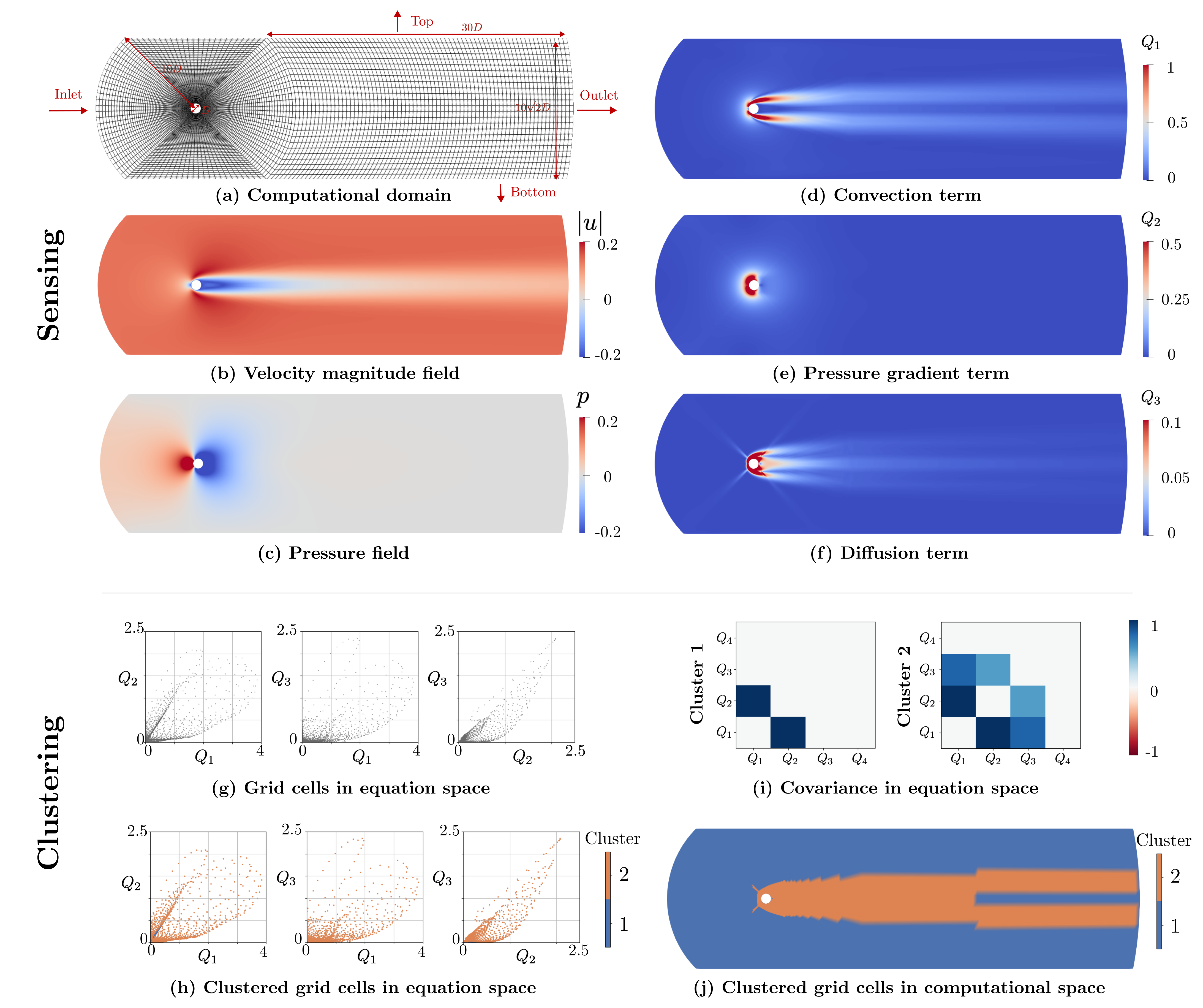}
			\caption{An illustration of (a) the computational domain with grid and corresponding steady state solution via (b) velocity and (c) pressure fields at Re = 40. (d-f) Convection term ($Q_1$), pressure gradient term ($Q_2$) and diffusion term ($Q_3$). Interntial term ($Q_4$) is not shown here because the flow is steady and $Q_4 = 0$. Distribution of grid cells in the equation space (g) without clustering and (h) with clustering and (i) corresponding covariance of GMM in the equation space. (j) Clustered/tagged grid cells in the computational space.}
			\label{fig:DBA}
		\end{figure}

		First we consider the numerical simulation of cylinder flow at $Re = 40$. Figure \ref{fig:DBA}(a) illustrates the computational domain and the distribution of grid cells in it. The 2D computational domain extends up to $47.74 D$ in the horizontal and $14.14 D$ in the vertical direction. The inlet is placed at $10 D$ upstream of the cylinder and a fixed momentum boundary condition is applied at this boundary using Dirichlet and Neumann conditions for the velocity and pressure fields, respectively. The outlet is placed at $37.74 D$ downstream of the cylinder and a free-shear boundary condition is applied using Neumann and Dirichlet conditions for the velocity and pressure fields, respectively. Top, and bottom boundaries are placed $5\sqrt{2} D$ above and below the cylinder and the same boundary conditions from the outlet are enforced. For the present case, we use a grid comprising 9075 cells, as shown in Figure \ref{fig:DBA}(a). The steady state solution fields $u$ and $p$ for this flow are shown in figures \ref{fig:DBA}(b) and (c), respectively.
		
		To illustrate the process of DBA for tagging the active grid cell, we consider the above obtained numerical solution fields $\bm{u}$ and $p$. As described in section \ref{sec:Method}, DBA requires computation of all the equation space terms $Q_i$s in the Navier stokes equation. The convection term ($Q_1$), pressure gradient ($Q_2$) and diffusion term ($Q_3$) are shown in figure \ref{fig:DBA}(d - f). It is observed that the convection and diffusion terms are dominant in the high shear regions of wake flow whereas the pressure gradient term mostly dominant the windward side of the cylinder where the flow accelerates rapidly due to high pressure gradient. It is noted here that due to steady nature of the flow, the solution fields $\bm{u},~p$ do not change in time and the inertial term ($Q_4$) is zero everywhere. Hence, for the application of DBA in this case, a three-dimensional equation space composed of only $Q_1,~Q_2$ and $Q_3$ can be considered, although the generalized nature of the algorithm also considers $Q_4$ in the calculation with no bearing on the final results as $Q_4 = 0$ in the present case.

		The active regions of the computational domain can be visually identified from the dominant structures formed by $Q_1,~Q_2$ and $Q_3$ in figures \ref{fig:DBA}(d - f). However, it is not trivial to quantitatively identify the corresponding grid cells for AMR, and the Gaussian mixture model (GMM) based unsupervised machine learning tool for clustering grid cells in the equation space is well suited to resolve this problem. The distribution of computational grid points in the equation space is shown in figures \ref{fig:DBA}(g). When a GMM with two components (one component each for active and passive grid cells) is trained on this distribution of grid cells in the equation space, the grid cells automatically get classified based on broad distribution of covariances in the $Q$-space. As shown in figure \ref{fig:DBA}(h), the grid cells are clustered such that cluster 1 (shown in blue color) contains all the grid cells where the convection term ($Q_1$) balance the pressure gradient term ($Q_2$) or has no interaction ($Q_1 = Q_2 = Q_3 = 0$) whereas cluster 2 (shown in orange color) contains the remaining grid cells which exhibit interaction among all three terms $Q_1$, $Q_2$ and $Q_3$. This pattern is quantitatively represented in figure \ref{fig:DBA}(i) through covariance of the GMM in the component-directions of equation space. It can be readily recognized that cluster 1 shows maximum interaction between $Q_1$ and $Q_2$ whereas cluster 2 shows interaction across all the components. 
		
		Figure \ref{fig:DBA}(j) illustrates the classification of grid cells in the computational domain across clusters 1 and 2. This clustering pattern is indicative of the fact that all the free-stream flow region is captured in cluster 1 where the flow can be considered inviscid and governed by the Bernoulli's principle. Hence, this steady inviscid flow regime is represented by the interaction between convection and pressure gradient terms which is also seen through the linear relationship followed by $Q_1$ and $Q_2$ for grid cells in cluster 1 (see figure \ref{fig:DBA}(h)). Cluster 2 captures the wake region of the flow which involves viscous interactions of the convection, pressure gradient and viscosity dominated flow features. With this interaction-based clustering of grid cells in the equation space, it is evident that cluster 2 represents the active region of the computational domain and DBA can efficiently tag grid cells for successive grid refinement in AMR.

		 To further demonstrate the dynamic adaptation of the grid tagging strategy across varying levels of grid refinement and evolving solutions over time, we consider unsteady flow over cylinder at $Re = 100$. 
		\begin{figure*}[ht]
			\centering
			\includegraphics[width=0.9\linewidth]{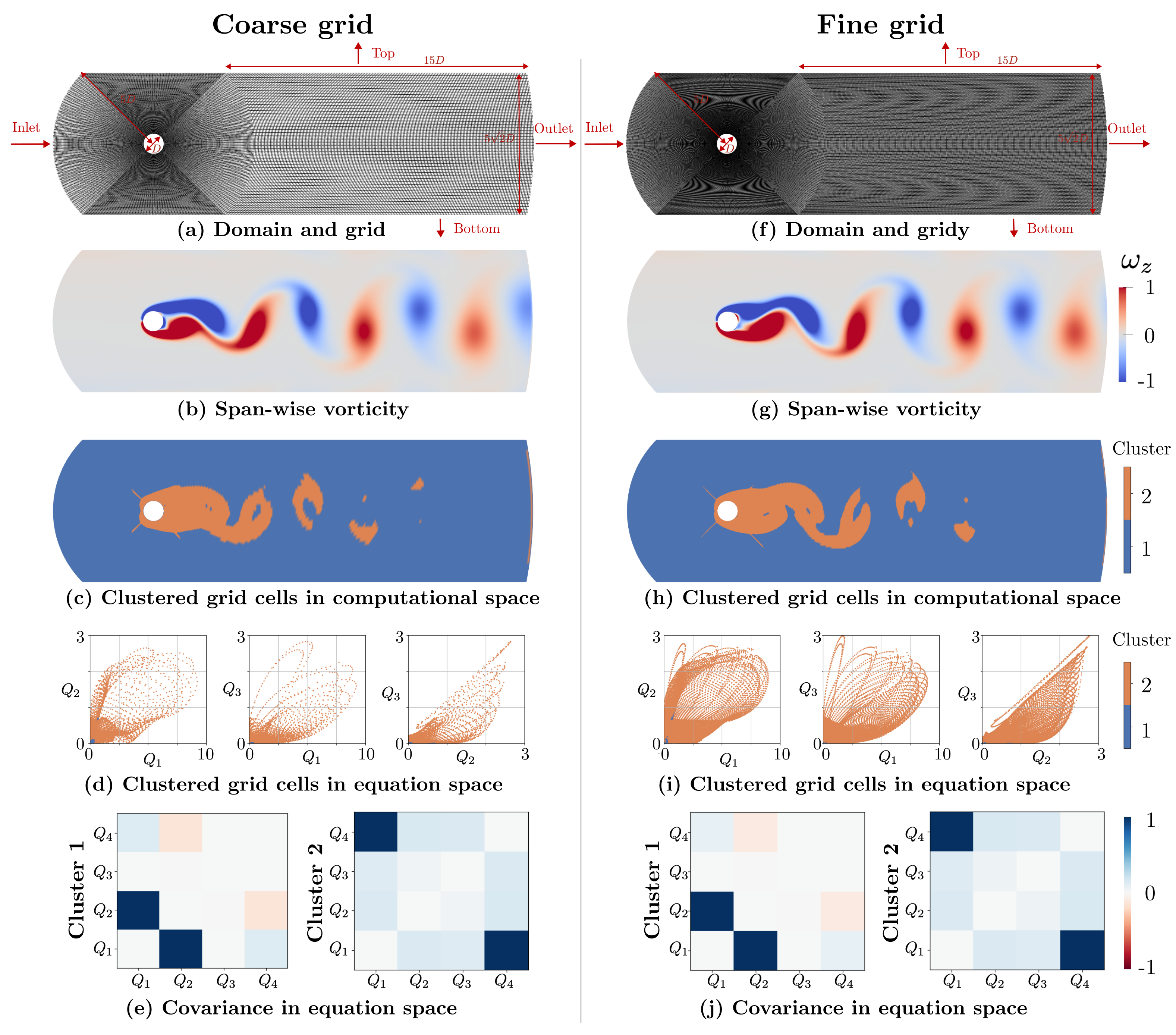}
			\caption{A comparison of DBA for an unsteady solution of flow past a cylinder at Re = 100 with coarse and fine grid. (a, f) Visualization of domain and grid. (b, g) Solution field at $t^* = 250$ represented through span-wise vorticity field. Clustered/tagged grid cells using GMM shown in (c, h) computational space and (d, i) equation space and (e, j) corresponding covariance in equation space.}
			\label{fig:Re100_grid_size_comp}
		\end{figure*}
		For the flow past a cylinder at $Re = 100$, the flow is characterized as 2D, laminar and unsteady. We utilize the previously described computational setup at $Re = 40$ with reduced computational domain size and refined grid, as illustrated in figure \ref{fig:Re100_grid_size_comp}(a, f), employing a coarse grid with 31703 cells and a fine grid with 130300 cells. The boundary conditions for the numerical simulation remains same as described earlier for $Re = 40$. Unsteady solutions are computed on both grids upto a 250 non-dimensional time units $(t^* = tU/D)$ where $U, D$ are the inlet flow velocity and diameter of the cylinder, respectively. Span-wise vorticity fields shown in figures \ref{fig:Re100_grid_size_comp}(b) and (g) illustrate the unsteady solution field obtained on the coarse and fine grids, respectively. It can be noted that the vortical structures simulated in the two cases are slighly shifted in phase due to different grid refinement although the solutions are plotted at the same time $t^* = 250$. Figures \ref{fig:Re100_grid_size_comp}(c) and (h) represent the clustering of coarse and fine grid cells, respectively in the computational space, and figures \ref{fig:Re100_grid_size_comp}(d) and (i) represent the clustering of coarse and fine grid cells, respectively in the equation space. Similar to the previous case at $Re = 40$, cluster 2 represents the grid cells with strong interaction in the equation space as demonstrated through the covariances observed across the two clusters in figures \ref{fig:Re100_grid_size_comp}(e) and (j). The subdomain, captured in cluster 2, encompasses the unsteady vortex shedding region downstream of the cylinder which is a well known phenomenon for cylinder flow at Re = 100. It is also noted that the relation between $Q_1$ and $Q_2$ is no longer linear because the flow is unsteady and the intertia term ($Q_4$) also contribute to the balance equation in the freestream flow. 
		
		From examining the cluster patterns in the computational space, it becomes evident that the area covered by cluster 2 on the fine grid is approximately same as the area covered by the same cluster on the coarse grid. Through comparison of the tagged grid cells on the coarse and fine grids, it is also noted that the spatial locations of the tagged cells overlap with strong vortical wake region of the corresponding solution as shown in figures  \ref{fig:Re100_grid_size_comp}(b) and (g). This illustrates the fact that DBA can identify active grid cells for refinement on a temporally evolving solution field, however, a more suitable use of the trained GMM will be to train the model once on a coarse grid and reuse it on temporally evolving (dynamically refined) grid over future time steps. 
		
\begin{figure*}[ht]
			\centering
			\includegraphics[width=0.9\linewidth]{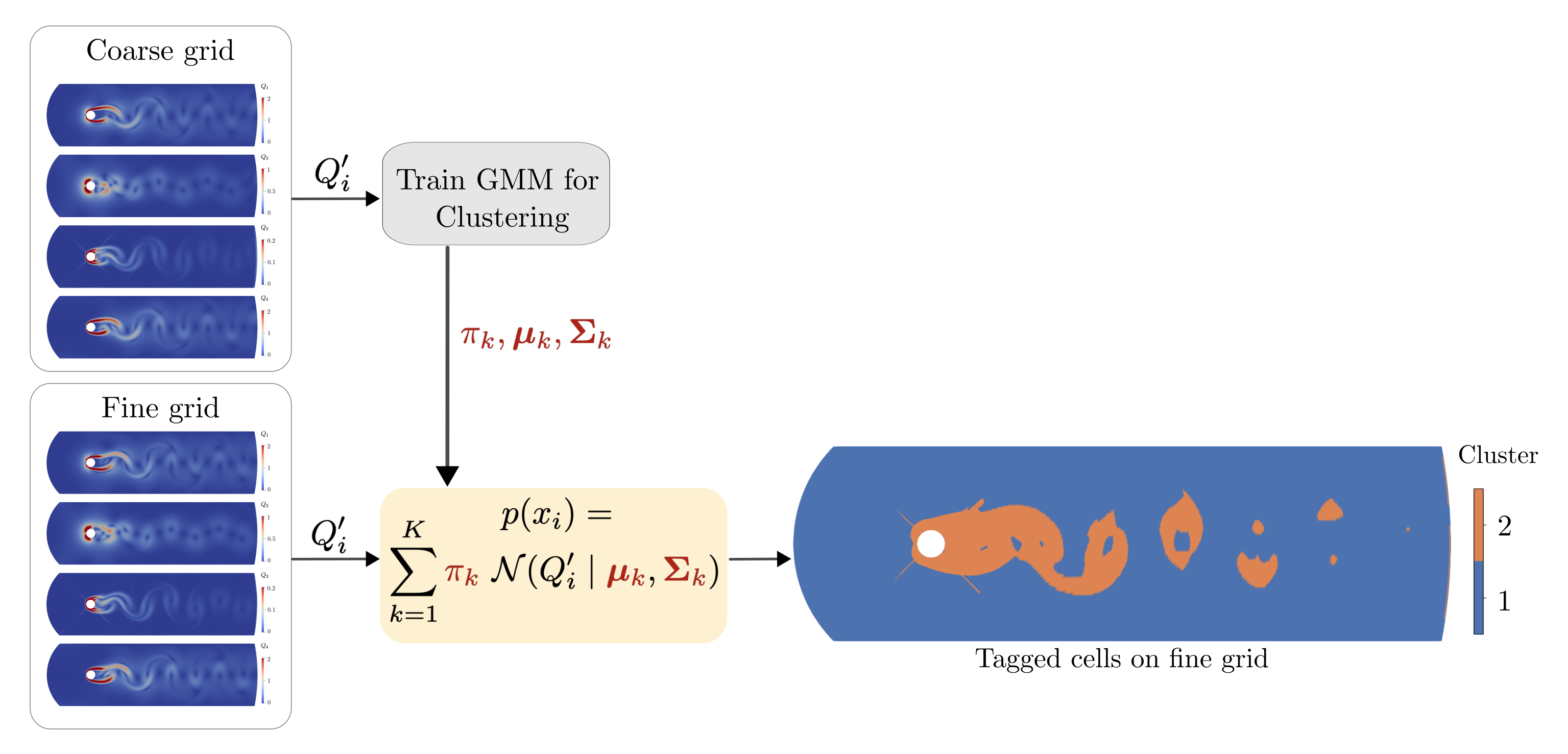}
			\caption{A schematic to show the application of GMM model on a fine grid grid that is trained through DBA on a coarse grid.}
			\label{fig:Re100_grid_mapping}
		\end{figure*}

		To explore the above prospect of GMM model for grid tagging, we train a GMM on the coarse grid and reuse this model on the solution ($\bm{u},~p$) obtained on a fine grid. This process is schematically shown in figure \ref{fig:Re100_grid_mapping}. From comparison of tagged region in figure \ref{fig:Re100_grid_mapping} with the tagged region in figure \ref{fig:Re100_grid_size_comp}(h), It is observed that the two are very similar, suggesting that the GMM model effectively encodes the interaction strength in the equation space, accounting for dependencies on space, time, and grid resolution. The model's performance is further demonstrated in the next section through adaptively refined simulations of 3D turbulent flow over a cylinder.

		\section{Results}\label{sec:Results and analysis}

\begin{figure*}[ht]
			\centering
			\includegraphics[width=0.9\linewidth]{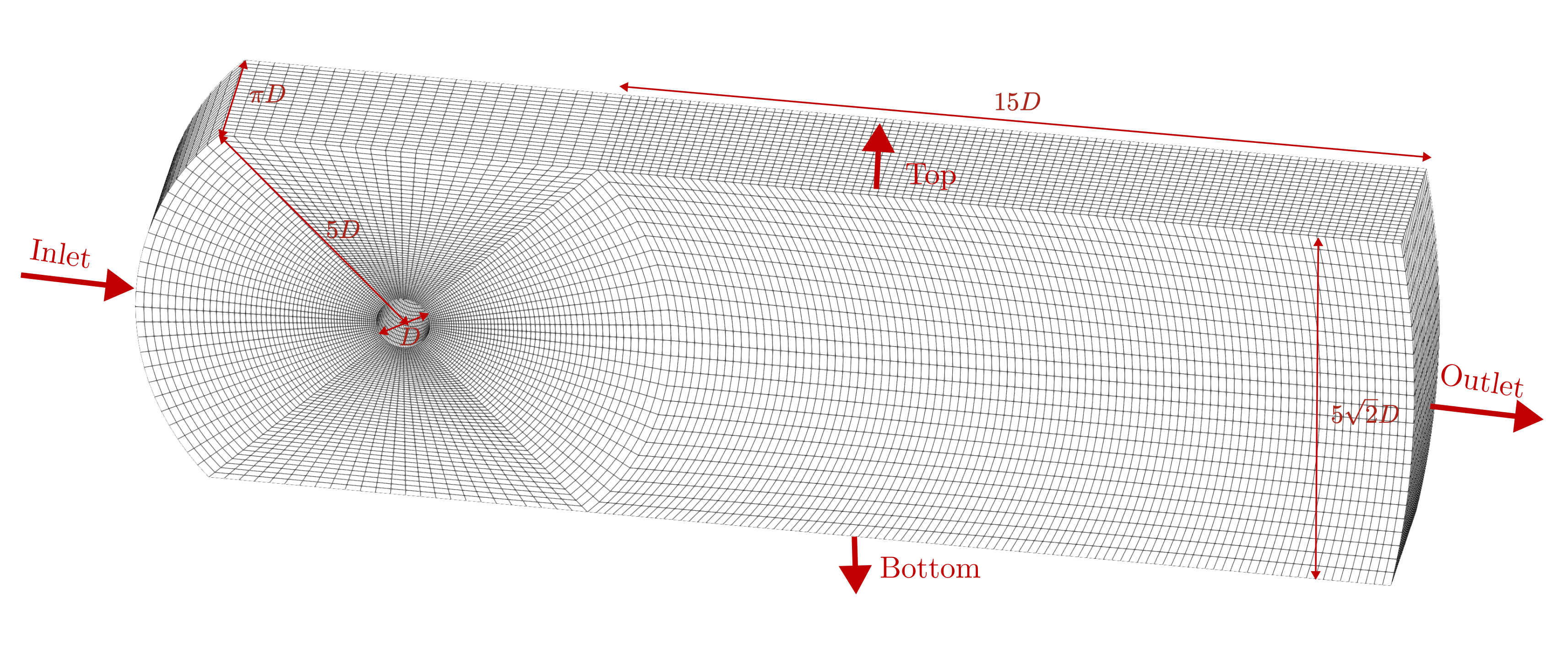}
			\caption{An illustration of 3D computational domain with coarse grid for large-eddy simulations.}
			\label{fig:domain}
		\end{figure*}

		The three-dimensional large eddy simulation (LES) of unsteady turbulent flow past a circular cylinder at subcritical Reynolds number $Re_D = 3900$ is considered as a test case to demonstrate the application of DBA based AMR strategy. To assess the accuracy and efficiency of AMR, a set of baseline simulations are performed on three levels of grid refinments. A 3D computational domain as shown in figure \ref{fig:domain} is generated by extruding the 2D setup described earlier. The same boundary conditions are adopted here and additional two boundaries in the spanwise direction are modelled to be periodic. Subgrid stresses from turbulence in LES is modelled using wall-adapting local eddy-viscosity (WALE) model \cite{nicoud1999subgrid}. For LES computation, the instantaneous velocity and pressure fields in Eq. \ref{eq:governingEqn} are replaced by the filtered velocity and pressure fields. Additionally $Re$ in $Q_3$ can be redefined as $Re_{eff} = UD/\nu_\textup{eff}$ where $\nu_\textup{eff}$ is the eddy viscosity calculated from WALE model. The distribution of cell sizes for the coarse grid in the domain is shown in figure \ref{fig:domain}. Grid-1 contains a total of approximately 170000 grid cells. Grid-2 (~1.3 million cells) and Grid-3 (~10.6 million cells) are obtained from uniformly refining coarse grid cells ($\Delta$) in Grid-1 by $1/2$ and $1/4$, respectively.
		
\begin{figure*}[ht]
			\centering
			\includegraphics[width=\linewidth]{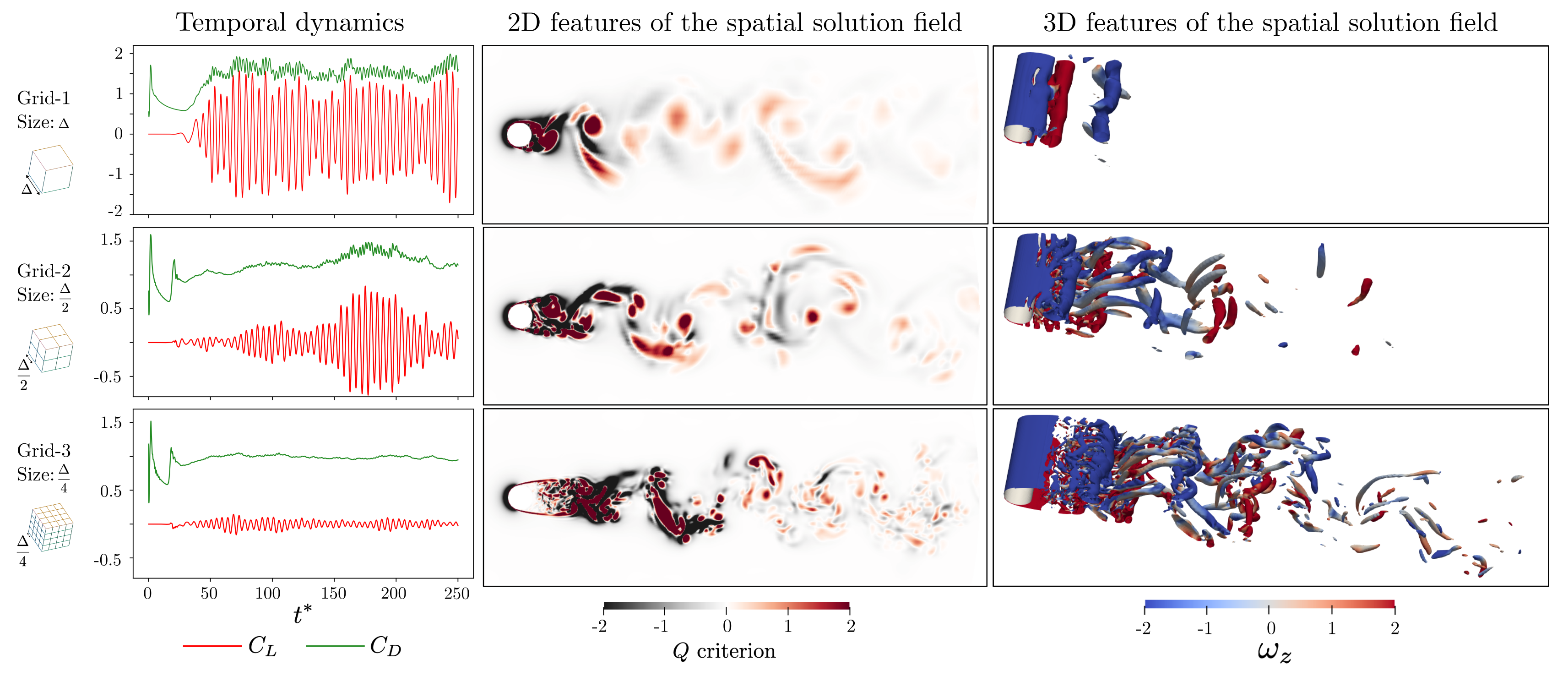}
			\caption{Dependence of spatio-temporal solution of flow past a cylinder at Re = 3900 on grid resolution. Top, middle and bottom rows of panels represent solution on Grid-1 (~$1.7\times 10^5$ cells), Grid-2 (~$1.3\times 10^6$ cells) and Grid-3 (~$1.06\times 10^7$ cells). Left column shows temporal dynamics of the solution through unsteady variation in force coefficients ($C_L$ and $C_D$). Middle column represents the 2D spatial features of solution via $Q$-criterion at $t^* = 250$. Right column represents the iso-contours of $Q$-criterion = 2 at $t^*=250$ colored with values of span-wise component of vorticity field ($\omega_z$).}
			\label{fig:gridConvergence}
		\end{figure*}

		The unsteady flow solutions obtained on the three levels of grid refinement are summarized in figure \ref{fig:gridConvergence}. The change in temporal dynamics of the solution field with grid refinment is shown through variation in unsteady lift and drag coefficients ($C_L$ and $C_D$). It is observed that the mean drag coefficient and variance in both lift and drag coefficient decreases with better grid refinement. This trend is further accompanied by changes in spatial flow features in the turbulent near-wake region. With successive grid refinement, broader range of length scales are captured from wider range of resolved turbulent eddies in the flow simulation. 
		
\begin{figure*}[h]
			\centering
			\includegraphics[width=0.85\linewidth]{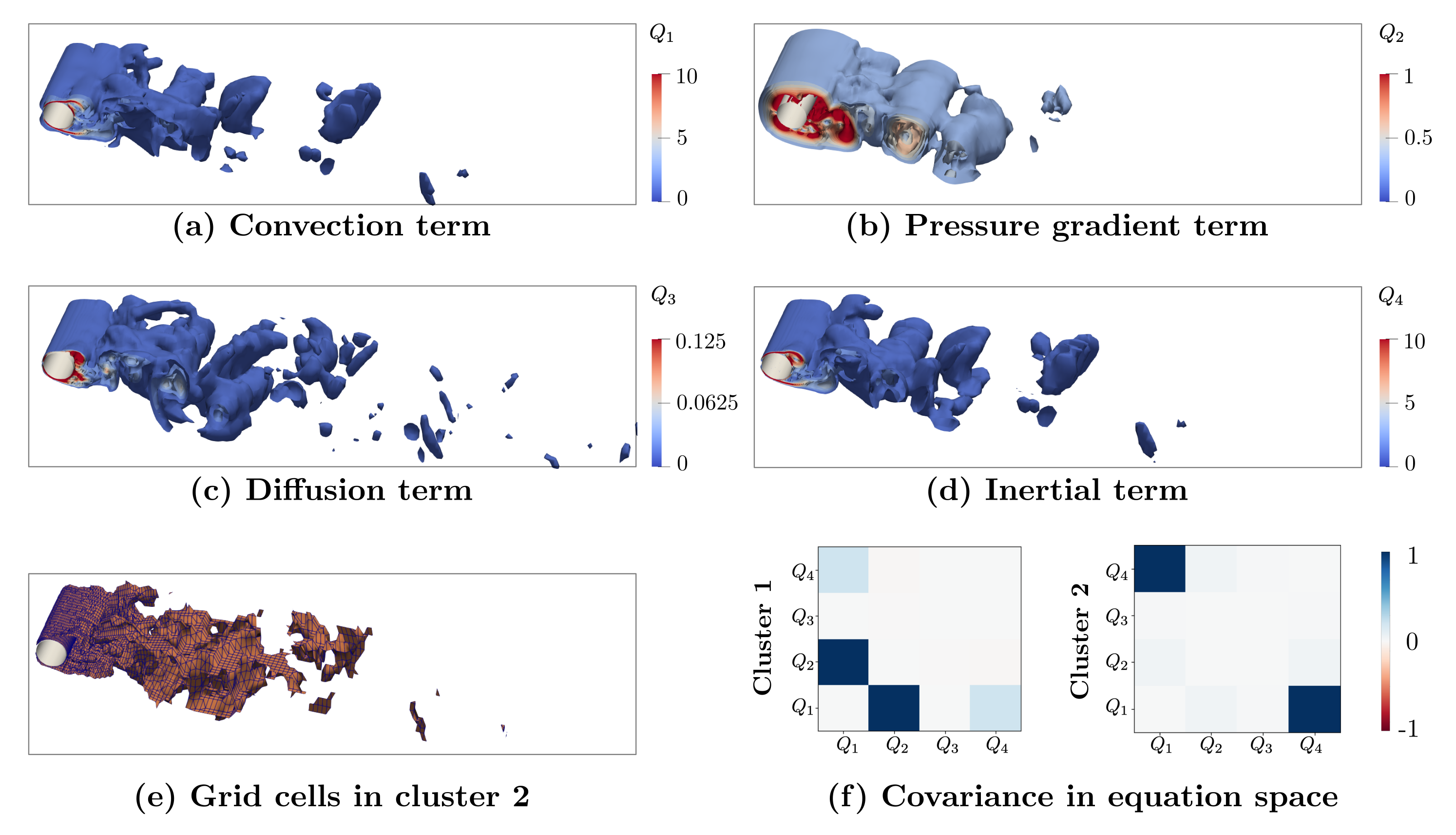}
			\caption{(a-d) Iso-contours of terms in the equation space $Q_is$. (e) Active grid cells (cluster 2) tagged through DBA. (f) Covariance in the equation space obtained from the trained GMM.}
			\label{fig:Re3900_DBA}
		\end{figure*}

		To employ AMR in the present setup, we use Grid-1 as the level 0 grid with cell size represented by $\Delta$. Maximum 2 levels of grid cell refinement is allowed at any location in AMR simulations which ensures that the maximum grid resolution at any spatial location is $\Delta/4$ and a comparison can be made with solution obtained from Grid-3 with resolution $\Delta/4$. The AMR simulation is restarted from non-dimensional solution time $t^* = 100$. To apply DBA on the intial solution field, equation space terms $Q_1,~Q_2,~Q_3,~Q_4$ are calculated as shown in figures \ref{fig:Re3900_DBA} (a-d). From visualization of the $Q_is$, it can be qualitatively observed that the active region of the domain comprises the wake region. To quantitatively tag the grid cells in this region, a GMM model is trained once at $t^* = 100$ using $Q_is$ obtained at Grid-1 employing the method shown earlier. The tagged grid cells at $t^*=100$ and the covariance matrix in the equation space are shown in figures \ref{fig:Re3900_DBA}(e) and \ref{fig:Re3900_DBA}(f). In adaptively refined simulation, the tagged grid cells are refined using OpenFOAM's dynamic mesh library and the initially trained GMM is reused to tag grid cells at future timesteps based on temporally evolving solution fields. 
		
		\begin{figure*}[ht]
			\centering
			\includegraphics[width=0.85\linewidth]{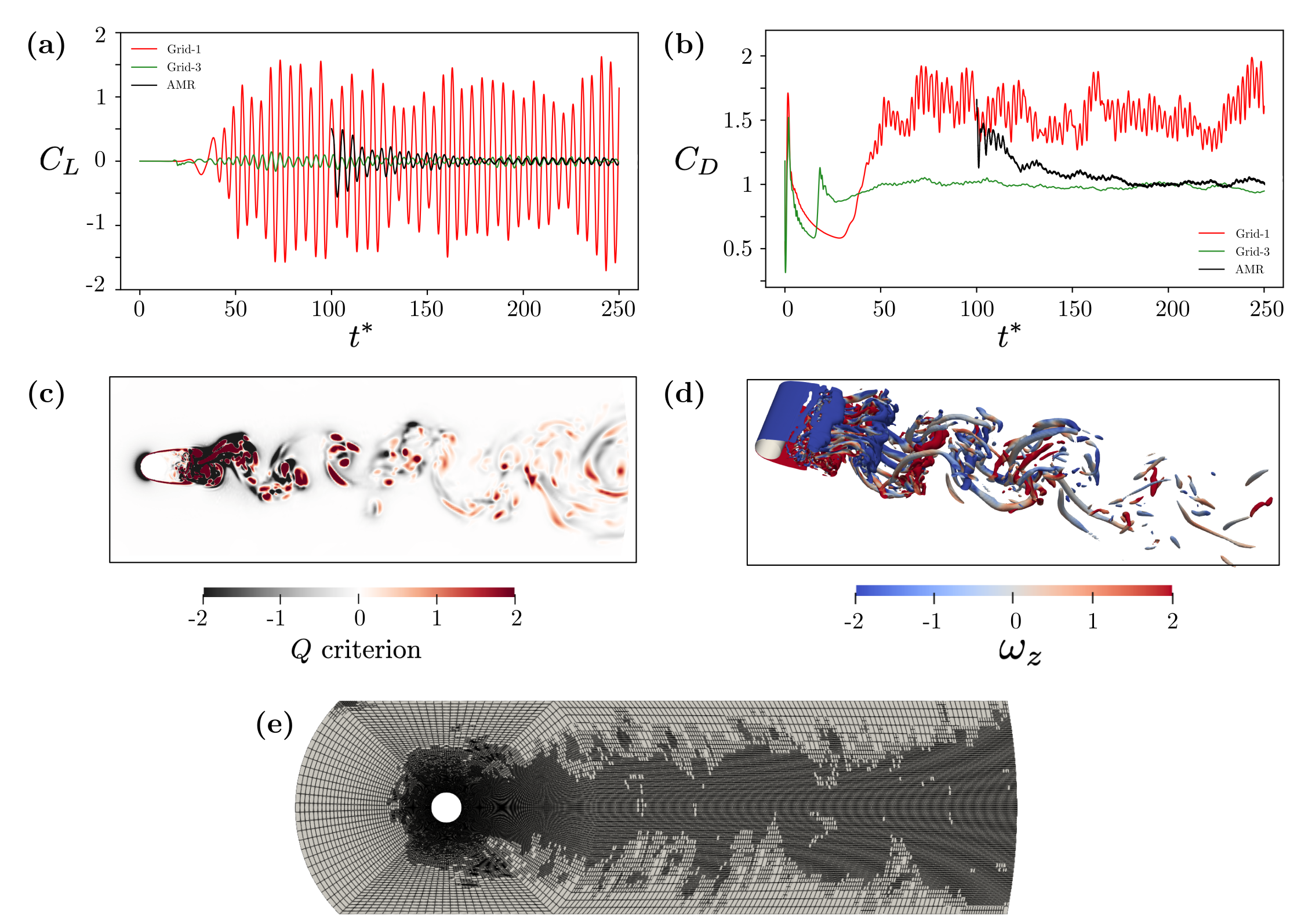}
			\caption{(a, b) Comparison of force coefficients ($C_L$ and $C_D$) from DBA based adaptively refined simulation on Grid-1 and Grid-3. (c) 2D spatial features of adaptively refined solution via $Q$-criterion at $t^* = 250$. (d) Iso-contours of $Q$-criterion = 2 at $t^*=250$ colored with values of span-wise component of vorticity field ($\omega_z$). (e) Adaptively refined grid representing distribution of cells in the domain after transient flow has passed ($t^* > 180$).}
			\label{fig:performance_V0}
		\end{figure*}
		
		\begin{figure*}[ht]
		\centering
		\includegraphics[width=0.9\linewidth]{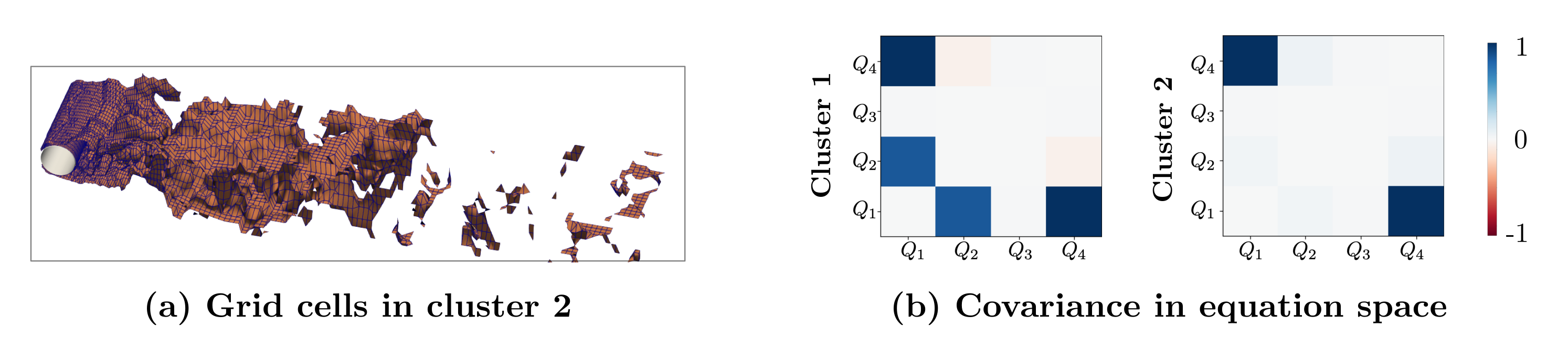}
		\caption{(a) Active grid cells (cluster 2) tagged through modified DBA. (b) Covariance in the equation space obtained from the trained GMM.}
		\label{fig:Re3900_DBA_1by3}
		\end{figure*}
		
		The time history of lift and drag coefficients computed on an adaptively refined grid are compared with the time history computed on Grid-1 and Grid-3 as shown in figures \ref{fig:performance_V0}(a) and (b). It can be easily seen that the force coefficients settle down close to the unsteady values observed on Grid-3 after the initial transient around $t^*\approx180$. Comparing the 2D and 3D spatial features of the solution field on the adaptively refined grid in Figures \ref{fig:performance_V0}(c) and (d) with those on Grid-3 in Figure \ref{fig:gridConvergence}, it is evident that the AMR strategy effectively resolves wake flow structures, which play a significant role in accurately predicting the forces on the cylinder. The AMR approach avoids refining the grid in the freestream region, where flow structures are less influential on force predictions. Figure \ref{fig:performance_V0}(e) shows a 2D projection of the computational grid, representing the distribution of adaptively refined cells in the domain after the transient flow phase ($t^* > 180$). The total cell count in the adaptively refined simulation stabilizes around 6.28 million, approximately 40\% less than the cell count on Grid-3 (~10.6 million cells). However, some grid cells are spuriously tagged in the freestream region due to numerical errors and local variations in grid size.
		
		To reduce the dependence of DBA-based grid tagging on local variations in grid size, we introduce a modified DBA where the $Q_i$ terms are weighted by the local grid size. In this modified analysis, instead of training the GMM on $Q_i$, we consider $Q_i \times \Delta$, where $\Delta$ is calculated as the cube root of the local grid cell volume ($\Delta V$). This strategy prioritizes coarser grid cells for further refinement over those that are already refined. The tagged grid cells at $t^* = 100$ and the covariance matrix in the equation space for the modified DBA are shown in Figure \ref{fig:Re3900_DBA_1by3}, which are very similar to those observed in the original DBA analysis in Figures \ref{fig:Re3900_DBA}(e) and (f).

\begin{figure*}[ht]
			\centering
			\includegraphics[width=\linewidth]{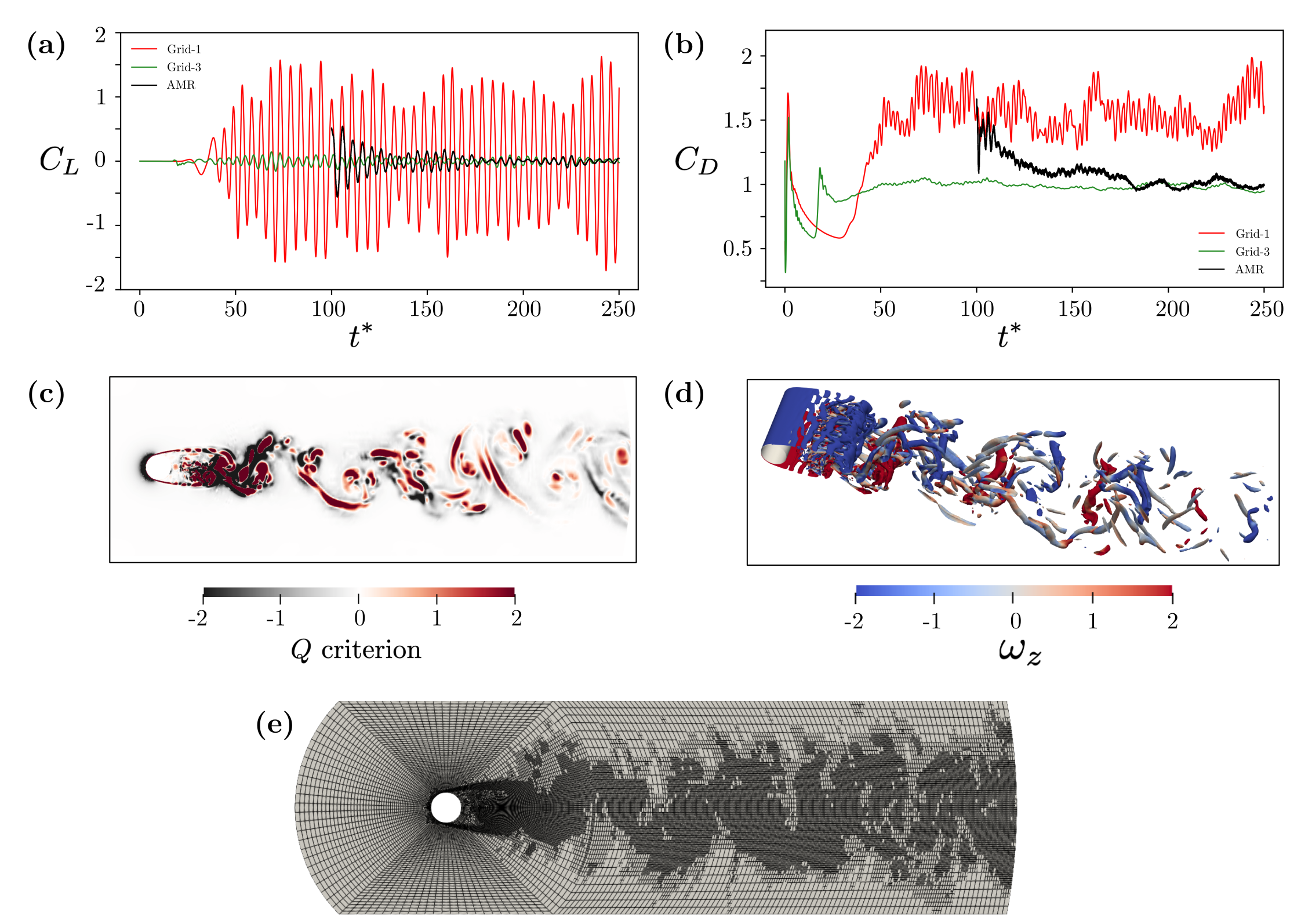}
			\caption{(a, b) Comparison of force coefficients ($C_L$ and $C_D$) from modified DBA based adaptively refined simulation on Grid-1 and Grid-3. (c) 2D spatial features of adaptively refined solution via $Q$-criterion at $t^* = 250$. (d) Iso-contours of $Q$-criterion = 2 at $t^*=250$ colored with values of span-wise component of vorticity field ($\omega_z$). (e) Adaptively refined grid representing distribution of cells in the domain after transient flow has passed ($t^* > 180$).}
			\label{fig:performance_V1by3}
		\end{figure*}

		The improved performance of AMR with modified DBA in the adaptively refined grid simulation is illustrated in Figure \ref{fig:performance_V1by3}. A comparison of the time history of force coefficients and the 2D and 3D flow structures, shown using the Q-criterion in Figures \ref{fig:performance_V1by3}(a-d) for AMR with modified DBA and \ref{fig:performance_V0}(a-d) for AMR with the original DBA, reveals that the solution accuracy is approximately the same. However, the number of grid cells in the AMR simulation with modified DBA is 2.77 million, which is 56\% lower than in the AMR simulation with original DBA (6.28 million) and 74\% lower than in Grid-3. Comparing the grid distribution in the adaptively refined simulations with the original and modified DBA, as shown in Figures \ref{fig:performance_V0}(e) and \ref{fig:performance_V1by3}(e), respectively, further indicates that the refined grid cells are significantly fewer in the modified DBA case. Additionally, the refined cells are more closely aligned with the high-vorticity and shear regions observed in the flow.

\begin{figure*}[ht]
			\centering
			\includegraphics[width=\linewidth]{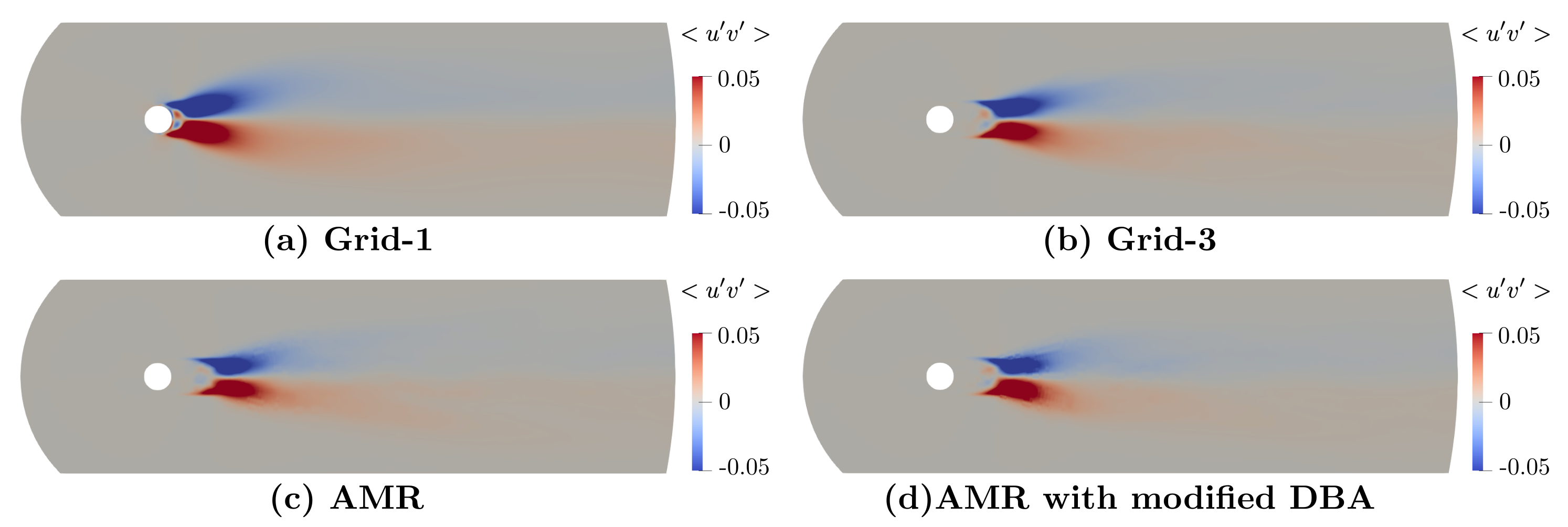}
			\caption{Comparison of Reynolds stress component $<u'v'>$ across turbulent flow solutions on (a) Grid-1 (~$1.7\times 10^5$ cells), (b) Grid-3 (~$1.06\times 10^7$ cells), (c) DBA based adaptively refined grid (~$6.28\times 10^6$ cells) and (d) modified DBA based adaptively refined grid (~$2.77\times 10^6$ cells).}
			\label{fig:uv}
		\end{figure*}

		To further substantiate the efficiency of the AMR strategy with both the original and modified DBA, we investigate the resolved Reynolds stresses in the turbulent flow. To compute Reynolds stresses, the solution obtained in the LES simulation is time-averaged over 300 time units after the flow becomes statistically stationary ($t^* > 180$). In Figure \ref{fig:uv}, the Reynolds stress component $\langle u'v' \rangle$ is compared for uniformly refined grids (Grid-1 and Grid-3) and adaptively refined grids with both the original and modified DBA. A significant difference is observed in $\langle u'v' \rangle$ between Grid-1 and Grid-3 due to the large difference in grid resolution, while the adaptively refined simulations with both DBA versions produce $\langle u'v' \rangle$ values comparable to those on the refined Grid-3. Notably, this comparable accuracy is achieved in the AMR simulations with the original and modified DBA while reducing the number of grid cells by 40\% and 70\%, respectively, compared to Grid-3.

		\begin{figure}[ht]
			\centering
			\includegraphics[width=0.6\linewidth]{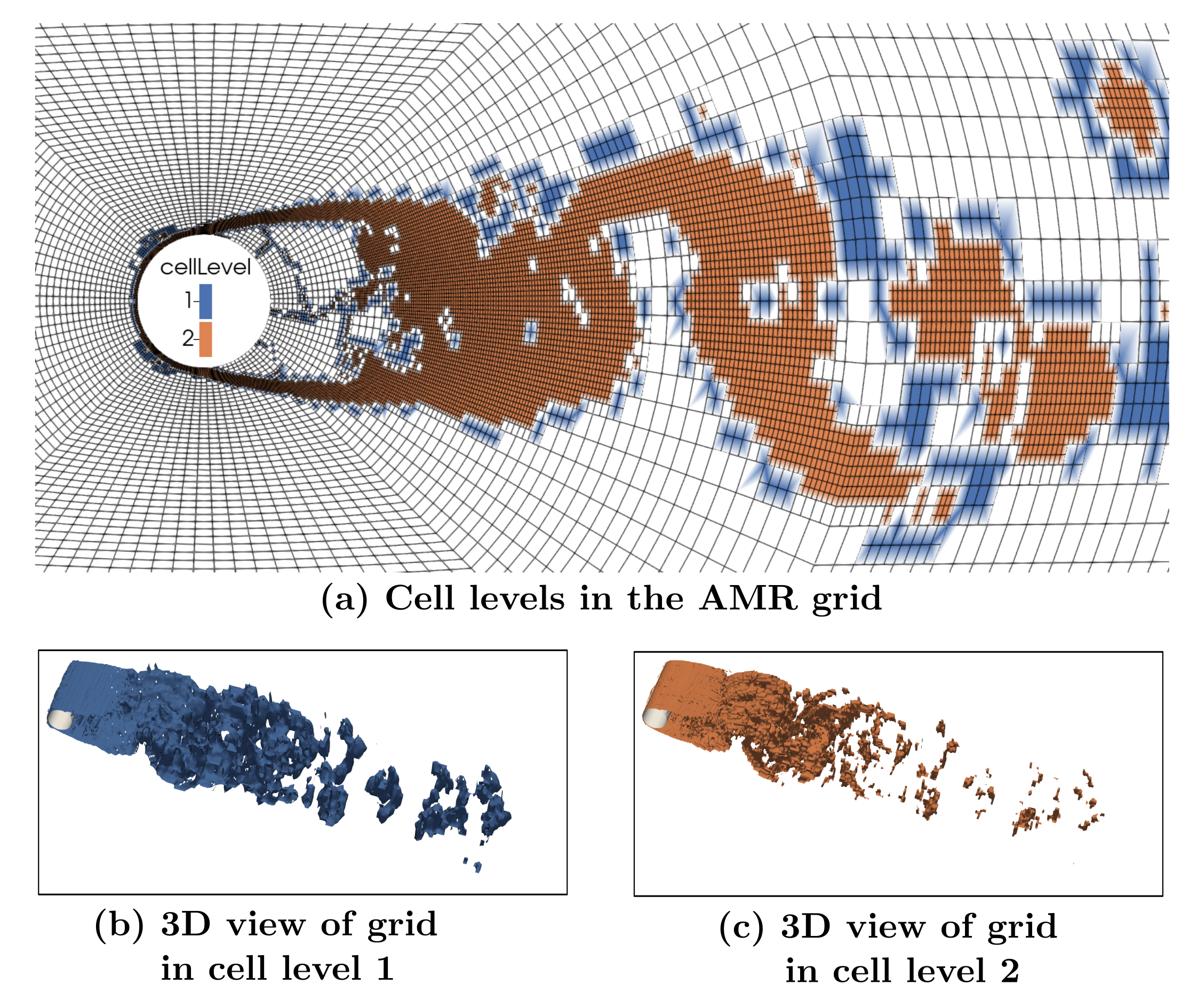}
			\caption{A 2D and 3D visualization of the modified DBA based adaptively refined grid cells in the simulation of flow past a cylinder at Re = 3900.}
			\label{fig:AMR_grid}
		\end{figure}
		
		A close inspection of the adaptively refined grid cells with modified DBA at levels 1 and 2, as shown in Figure \ref{fig:AMR_grid}(a), reveals that the AMR strategy accurately resolves the attached boundary layer, separated shear layer, and turbulent eddies in the near-wake region of the flow. While level 2 grid cells capture most of the details in these regions, level 1 grid cells primarily occupy the interface between level 0 (unrefined) and level 2 (refined) grid cells. It is noteworthy that throughout this adaptive refinement process, no heuristic-based sensors or predefined optimization outputs were used. Additionally, no user input requiring a priori knowledge of the solution was provided.

		
\section{Conclusion}\label{sec:Conclusion}

This work presents an adaptive mesh refinement (AMR) strategy that leverages dominant balance analysis (DBA) to automatically identify regions of high interaction within the computational domain, achieving efficient grid tagging and refinement for complex partial differential equations (PDEs). By employing a Gaussian mixture model (GMM) in the equation space, this method accurately and efficiently classifies grid cells into active and passive regions without the need for heuristic-based sensors or a priori user input. Applied to the Navier-Stokes equations, a highly non-linear and computationally demanding PDE, this AMR strategy demonstrates robust performance across both steady and unsteady flows, maintaining solution accuracy while significantly reducing computational costs. The effectiveness of the proposed method is validated through extensive simulations of flow around a cylinder, where it captures essential flow dynamics and force predictions while reducing grid cell counts by up to 70\% compared to traditional uniformly refined grids.

The modular, problem-independent design of this AMR strategy makes it adaptable for various computationally intensive problems in CFD and beyond. Ongoing work is focused on extending the method to applications involving fluid-structure interaction and compressible flows, enhancing its scalability and flexibility. The proposed DBA-based AMR method addresses existing challenges in adaptive meshing by offering a practical and efficient approach for high-fidelity simulations. This work contributes to the development of more autonomous and computationally efficient AMR strategies, making them accessible for simulations that demand both high accuracy and resource efficiency.

		\section*{Funding Sources}
		
		AGN acknowledges the support from the Department of Energy Early Career Research Award (Award no: DE-SC0022945, PM: Dr. William Spotz).
		
		\nocite{*}
		\bibliography{aipsamp}

\begin{thebibliography}{36}%
\makeatletter
\providecommand \@ifxundefined [1]{%
 \@ifx{#1\undefined}
}%
\providecommand \@ifnum [1]{%
 \ifnum #1\expandafter \@firstoftwo
 \else \expandafter \@secondoftwo
 \fi
}%
\providecommand \@ifx [1]{%
 \ifx #1\expandafter \@firstoftwo
 \else \expandafter \@secondoftwo
 \fi
}%
\providecommand \natexlab [1]{#1}%
\providecommand \enquote  [1]{``#1''}%
\providecommand \bibnamefont  [1]{#1}%
\providecommand \bibfnamefont [1]{#1}%
\providecommand \citenamefont [1]{#1}%
\providecommand \href@noop [0]{\@secondoftwo}%
\providecommand \href [0]{\begingroup \@sanitize@url \@href}%
\providecommand \@href[1]{\@@startlink{#1}\@@href}%
\providecommand \@@href[1]{\endgroup#1\@@endlink}%
\providecommand \@sanitize@url [0]{\catcode `\\12\catcode `\$12\catcode
  `\&12\catcode `\#12\catcode `\^12\catcode `\_12\catcode `\%12\relax}%
\providecommand \@@startlink[1]{}%
\providecommand \@@endlink[0]{}%
\providecommand \url  [0]{\begingroup\@sanitize@url \@url }%
\providecommand \@url [1]{\endgroup\@href {#1}{\urlprefix }}%
\providecommand \urlprefix  [0]{URL }%
\providecommand \Eprint [0]{\href }%
\providecommand \doibase [0]{http://dx.doi.org/}%
\providecommand \selectlanguage [0]{\@gobble}%
\providecommand \bibinfo  [0]{\@secondoftwo}%
\providecommand \bibfield  [0]{\@secondoftwo}%
\providecommand \translation [1]{[#1]}%
\providecommand \BibitemOpen [0]{}%
\providecommand \bibitemStop [0]{}%
\providecommand \bibitemNoStop [0]{.\EOS\space}%
\providecommand \EOS [0]{\spacefactor3000\relax}%
\providecommand \BibitemShut  [1]{\csname bibitem#1\endcsname}%
\let\auto@bib@innerbib\@empty
\bibitem [{\citenamefont {Buljac}\ \emph {et~al.}(2016)\citenamefont {Buljac},
  \citenamefont {D{\v{z}}ijan}, \citenamefont {Korade}, \citenamefont
  {Krizmani{\'c}},\ and\ \citenamefont {Kozmar}}]{buljac2016automobile}%
  \BibitemOpen
  \bibfield  {author} {\bibinfo {author} {\bibfnamefont {A.}~\bibnamefont
  {Buljac}}, \bibinfo {author} {\bibfnamefont {I.}~\bibnamefont
  {D{\v{z}}ijan}}, \bibinfo {author} {\bibfnamefont {I.}~\bibnamefont
  {Korade}}, \bibinfo {author} {\bibfnamefont {S.}~\bibnamefont
  {Krizmani{\'c}}}, \ and\ \bibinfo {author} {\bibfnamefont {H.}~\bibnamefont
  {Kozmar}},\ }\bibfield  {title} {\enquote {\bibinfo {title} {Automobile
  aerodynamics influenced by airfoil-shaped rear wing},}\ }\href@noop {}
  {\bibfield  {journal} {\bibinfo  {journal} {International journal of
  automotive technology}\ }\textbf {\bibinfo {volume} {17}},\ \bibinfo {pages}
  {377--385} (\bibinfo {year} {2016})}\BibitemShut {NoStop}%
\bibitem [{\citenamefont {Rizzi}\ and\ \citenamefont
  {Luckring}(2021)}]{rizzi2021historical}%
  \BibitemOpen
  \bibfield  {author} {\bibinfo {author} {\bibfnamefont {A.}~\bibnamefont
  {Rizzi}}\ and\ \bibinfo {author} {\bibfnamefont {J.~M.}\ \bibnamefont
  {Luckring}},\ }\bibfield  {title} {\enquote {\bibinfo {title} {Historical
  development and use of cfd for separated flow simulations relevant to
  military aircraft},}\ }\href@noop {} {\bibfield  {journal} {\bibinfo
  {journal} {Aerospace Science and Technology}\ }\textbf {\bibinfo {volume}
  {117}},\ \bibinfo {pages} {106940} (\bibinfo {year} {2021})}\BibitemShut
  {NoStop}%
\bibitem [{\citenamefont {Crain}\ and\ \citenamefont {van~de
  Voort}(2023)}]{crain2023hydrodynamical}%
  \BibitemOpen
  \bibfield  {author} {\bibinfo {author} {\bibfnamefont {R.~A.}\ \bibnamefont
  {Crain}}\ and\ \bibinfo {author} {\bibfnamefont {F.}~\bibnamefont {van~de
  Voort}},\ }\bibfield  {title} {\enquote {\bibinfo {title} {Hydrodynamical
  simulations of the galaxy population: enduring successes and outstanding
  challenges},}\ }\href@noop {} {\bibfield  {journal} {\bibinfo  {journal}
  {Annual Review of Astronomy and Astrophysics}\ }\textbf {\bibinfo {volume}
  {61}},\ \bibinfo {pages} {473--515} (\bibinfo {year} {2023})}\BibitemShut
  {NoStop}%
\bibitem [{\citenamefont {Lin-Lin}, \citenamefont {Hui},\ and\ \citenamefont
  {Chui-Jie}(2016)}]{lin2016three}%
  \BibitemOpen
  \bibfield  {author} {\bibinfo {author} {\bibfnamefont {Z.}~\bibnamefont
  {Lin-Lin}}, \bibinfo {author} {\bibfnamefont {G.}~\bibnamefont {Hui}}, \ and\
  \bibinfo {author} {\bibfnamefont {W.}~\bibnamefont {Chui-Jie}},\ }\bibfield
  {title} {\enquote {\bibinfo {title} {Three-dimensional numerical simulation
  of a bird model in unsteady flight},}\ }\href@noop {} {\bibfield  {journal}
  {\bibinfo  {journal} {Computational Mechanics}\ }\textbf {\bibinfo {volume}
  {58}},\ \bibinfo {pages} {1--11} (\bibinfo {year} {2016})}\BibitemShut
  {NoStop}%
\bibitem [{\citenamefont {Macias}\ \emph {et~al.}(2020)\citenamefont {Macias},
  \citenamefont {Souza}, \citenamefont {Junior},\ and\ \citenamefont
  {Oliveira}}]{macias2020three}%
  \BibitemOpen
  \bibfield  {author} {\bibinfo {author} {\bibfnamefont {M.~M.}\ \bibnamefont
  {Macias}}, \bibinfo {author} {\bibfnamefont {I.~F.}\ \bibnamefont {Souza}},
  \bibinfo {author} {\bibfnamefont {A.~C.~B.}\ \bibnamefont {Junior}}, \ and\
  \bibinfo {author} {\bibfnamefont {T.~F.}\ \bibnamefont {Oliveira}},\
  }\bibfield  {title} {\enquote {\bibinfo {title} {Three-dimensional viscous
  wake flow in fish swimming-a cfd study},}\ }\href@noop {} {\bibfield
  {journal} {\bibinfo  {journal} {Mechanics Research Communications}\ }\textbf
  {\bibinfo {volume} {107}},\ \bibinfo {pages} {103547} (\bibinfo {year}
  {2020})}\BibitemShut {NoStop}%
\bibitem [{\citenamefont {Tadamasa}\ and\ \citenamefont
  {Zangeneh}(2011)}]{tadamasa2011numerical}%
  \BibitemOpen
  \bibfield  {author} {\bibinfo {author} {\bibfnamefont {A.}~\bibnamefont
  {Tadamasa}}\ and\ \bibinfo {author} {\bibfnamefont {M.}~\bibnamefont
  {Zangeneh}},\ }\bibfield  {title} {\enquote {\bibinfo {title} {Numerical
  prediction of wind turbine noise},}\ }\href@noop {} {\bibfield  {journal}
  {\bibinfo  {journal} {Renewable energy}\ }\textbf {\bibinfo {volume} {36}},\
  \bibinfo {pages} {1902--1912} (\bibinfo {year} {2011})}\BibitemShut {NoStop}%
\bibitem [{\citenamefont {Yang}\ and\ \citenamefont
  {Griffin}(2021)}]{yang2021grid}%
  \BibitemOpen
  \bibfield  {author} {\bibinfo {author} {\bibfnamefont {X.~I.}\ \bibnamefont
  {Yang}}\ and\ \bibinfo {author} {\bibfnamefont {K.~P.}\ \bibnamefont
  {Griffin}},\ }\bibfield  {title} {\enquote {\bibinfo {title} {Grid-point and
  time-step requirements for direct numerical simulation and large-eddy
  simulation},}\ }\href@noop {} {\bibfield  {journal} {\bibinfo  {journal}
  {Physics of Fluids}\ }\textbf {\bibinfo {volume} {33}} (\bibinfo {year}
  {2021})}\BibitemShut {NoStop}%
\bibitem [{\citenamefont {Lucas}\ \emph {et~al.}(2014)\citenamefont {Lucas},
  \citenamefont {Ang}, \citenamefont {Bergman}, \citenamefont {Borkar},
  \citenamefont {Carlson}, \citenamefont {Carrington}, \citenamefont {Chiu},
  \citenamefont {Colwell}, \citenamefont {Dally}, \citenamefont {Dongarra}
  \emph {et~al.}}]{lucas2014doe}%
  \BibitemOpen
  \bibfield  {author} {\bibinfo {author} {\bibfnamefont {R.}~\bibnamefont
  {Lucas}}, \bibinfo {author} {\bibfnamefont {J.}~\bibnamefont {Ang}}, \bibinfo
  {author} {\bibfnamefont {K.}~\bibnamefont {Bergman}}, \bibinfo {author}
  {\bibfnamefont {S.}~\bibnamefont {Borkar}}, \bibinfo {author} {\bibfnamefont
  {W.}~\bibnamefont {Carlson}}, \bibinfo {author} {\bibfnamefont
  {L.}~\bibnamefont {Carrington}}, \bibinfo {author} {\bibfnamefont
  {G.}~\bibnamefont {Chiu}}, \bibinfo {author} {\bibfnamefont {R.}~\bibnamefont
  {Colwell}}, \bibinfo {author} {\bibfnamefont {W.}~\bibnamefont {Dally}},
  \bibinfo {author} {\bibfnamefont {J.}~\bibnamefont {Dongarra}},  \emph
  {et~al.},\ }\href@noop {} {\enquote {\bibinfo {title} {{D}{O}{E} advanced
  scientific computing advisory subcommittee ({ASCAC}) report: top ten exascale
  research challenges},}\ }\bibinfo {type} {Tech. Rep.}\ (\bibinfo
  {institution} {USDOE Office of Science (SC)(United States)},\ \bibinfo {year}
  {2014})\BibitemShut {NoStop}%
\bibitem [{\citenamefont {Berger}\ and\ \citenamefont
  {Oliger}(1984)}]{berger1984adaptive}%
  \BibitemOpen
  \bibfield  {author} {\bibinfo {author} {\bibfnamefont {M.~J.}\ \bibnamefont
  {Berger}}\ and\ \bibinfo {author} {\bibfnamefont {J.}~\bibnamefont
  {Oliger}},\ }\bibfield  {title} {\enquote {\bibinfo {title} {Adaptive mesh
  refinement for hyperbolic partial differential equations},}\ }\href@noop {}
  {\bibfield  {journal} {\bibinfo  {journal} {Journal of computational
  Physics}\ }\textbf {\bibinfo {volume} {53}},\ \bibinfo {pages} {484--512}
  (\bibinfo {year} {1984})}\BibitemShut {NoStop}%
\bibitem [{\citenamefont {Freitas}(2002)}]{freitas2002issue}%
  \BibitemOpen
  \bibfield  {author} {\bibinfo {author} {\bibfnamefont {C.~J.}\ \bibnamefont
  {Freitas}},\ }\bibfield  {title} {\enquote {\bibinfo {title} {The issue of
  numerical uncertainty},}\ }\href@noop {} {\bibfield  {journal} {\bibinfo
  {journal} {Applied Mathematical Modelling}\ }\textbf {\bibinfo {volume}
  {26}},\ \bibinfo {pages} {237--248} (\bibinfo {year} {2002})}\BibitemShut
  {NoStop}%
\bibitem [{\citenamefont {M{\"u}ller}\ and\ \citenamefont
  {Giles}(2001)}]{muller2001solution}%
  \BibitemOpen
  \bibfield  {author} {\bibinfo {author} {\bibfnamefont {J.-D.}\ \bibnamefont
  {M{\"u}ller}}\ and\ \bibinfo {author} {\bibfnamefont {M.}~\bibnamefont
  {Giles}},\ }\bibfield  {title} {\enquote {\bibinfo {title} {Solution adaptive
  mesh refinement using adjoint error analysis},}\ }in\ \href@noop {} {\emph
  {\bibinfo {booktitle} {15th AIAA Computational Fluid Dynamics Conference}}}\
  (\bibinfo {year} {2001})\ p.\ \bibinfo {pages} {2550}\BibitemShut {NoStop}%
\bibitem [{\citenamefont {Nemec}, \citenamefont {Aftosmis},\ and\ \citenamefont
  {Wintzer}(2008)}]{nemec2008adjoint}%
  \BibitemOpen
  \bibfield  {author} {\bibinfo {author} {\bibfnamefont {M.}~\bibnamefont
  {Nemec}}, \bibinfo {author} {\bibfnamefont {M.}~\bibnamefont {Aftosmis}}, \
  and\ \bibinfo {author} {\bibfnamefont {M.}~\bibnamefont {Wintzer}},\
  }\bibfield  {title} {\enquote {\bibinfo {title} {Adjoint-based adaptive mesh
  refinement for complex geometries},}\ }in\ \href@noop {} {\emph {\bibinfo
  {booktitle} {46th AIAA Aerospace Sciences Meeting and Exhibit}}}\ (\bibinfo
  {year} {2008})\ p.\ \bibinfo {pages} {725}\BibitemShut {NoStop}%
\bibitem [{\citenamefont {Li}, \citenamefont {Jameson},\ and\ \citenamefont
  {Allaneau}(2011)}]{li2011continuous}%
  \BibitemOpen
  \bibfield  {author} {\bibinfo {author} {\bibfnamefont {Y.}~\bibnamefont
  {Li}}, \bibinfo {author} {\bibfnamefont {A.}~\bibnamefont {Jameson}}, \ and\
  \bibinfo {author} {\bibfnamefont {Y.}~\bibnamefont {Allaneau}},\ }\bibfield
  {title} {\enquote {\bibinfo {title} {Continuous adjoint approach for adaptive
  mesh refinement},}\ }in\ \href@noop {} {\emph {\bibinfo {booktitle} {20th
  AIAA Computational Fluid Dynamics Conference}}}\ (\bibinfo {year} {2011})\
  p.\ \bibinfo {pages} {3982}\BibitemShut {NoStop}%
\bibitem [{\citenamefont {Hartmann}, \citenamefont {Held},\ and\ \citenamefont
  {Leicht}(2011)}]{hartmann2011adjoint}%
  \BibitemOpen
  \bibfield  {author} {\bibinfo {author} {\bibfnamefont {R.}~\bibnamefont
  {Hartmann}}, \bibinfo {author} {\bibfnamefont {J.}~\bibnamefont {Held}}, \
  and\ \bibinfo {author} {\bibfnamefont {T.}~\bibnamefont {Leicht}},\
  }\bibfield  {title} {\enquote {\bibinfo {title} {Adjoint-based error
  estimation and adaptive mesh refinement for the rans and k--$\omega$
  turbulence model equations},}\ }\href@noop {} {\bibfield  {journal} {\bibinfo
   {journal} {Journal of Computational Physics}\ }\textbf {\bibinfo {volume}
  {230}},\ \bibinfo {pages} {4268--4284} (\bibinfo {year} {2011})}\BibitemShut
  {NoStop}%
\bibitem [{\citenamefont {Fidkowski}\ and\ \citenamefont
  {Darmofal}(2011)}]{fidkowski2011review}%
  \BibitemOpen
  \bibfield  {author} {\bibinfo {author} {\bibfnamefont {K.~J.}\ \bibnamefont
  {Fidkowski}}\ and\ \bibinfo {author} {\bibfnamefont {D.~L.}\ \bibnamefont
  {Darmofal}},\ }\bibfield  {title} {\enquote {\bibinfo {title} {Review of
  output-based error estimation and mesh adaptation in computational fluid
  dynamics},}\ }\href@noop {} {\bibfield  {journal} {\bibinfo  {journal} {AIAA
  journal}\ }\textbf {\bibinfo {volume} {49}},\ \bibinfo {pages} {673--694}
  (\bibinfo {year} {2011})}\BibitemShut {NoStop}%
\bibitem [{\citenamefont {Hartmann}, \citenamefont {Meinke},\ and\
  \citenamefont {Schr{\"o}der}(2011)}]{hartmann2011level}%
  \BibitemOpen
  \bibfield  {author} {\bibinfo {author} {\bibfnamefont {D.}~\bibnamefont
  {Hartmann}}, \bibinfo {author} {\bibfnamefont {M.}~\bibnamefont {Meinke}}, \
  and\ \bibinfo {author} {\bibfnamefont {W.}~\bibnamefont {Schr{\"o}der}},\
  }\bibfield  {title} {\enquote {\bibinfo {title} {A level-set based
  adaptive-grid method for premixed combustion},}\ }\href@noop {} {\bibfield
  {journal} {\bibinfo  {journal} {Combustion and flame}\ }\textbf {\bibinfo
  {volume} {158}},\ \bibinfo {pages} {1318--1339} (\bibinfo {year}
  {2011})}\BibitemShut {NoStop}%
\bibitem [{\citenamefont {Aftosmis}\ and\ \citenamefont
  {Berger}(2002)}]{aftosmis2002multilevel}%
  \BibitemOpen
  \bibfield  {author} {\bibinfo {author} {\bibfnamefont {M.}~\bibnamefont
  {Aftosmis}}\ and\ \bibinfo {author} {\bibfnamefont {M.}~\bibnamefont
  {Berger}},\ }\bibfield  {title} {\enquote {\bibinfo {title} {Multilevel error
  estimation and adaptive h-refinement for cartesian meshes with embedded
  boundaries},}\ }in\ \href@noop {} {\emph {\bibinfo {booktitle} {40th AIAA
  Aerospace Sciences Meeting \& Exhibit}}}\ (\bibinfo {year} {2002})\ p.\
  \bibinfo {pages} {863}\BibitemShut {NoStop}%
\bibitem [{\citenamefont {Wu}\ \emph {et~al.}(1990)\citenamefont {Wu},
  \citenamefont {Zhu}, \citenamefont {Szmelter},\ and\ \citenamefont
  {Zienkiewicz}}]{wu1990error}%
  \BibitemOpen
  \bibfield  {author} {\bibinfo {author} {\bibfnamefont {J.}~\bibnamefont
  {Wu}}, \bibinfo {author} {\bibfnamefont {J.}~\bibnamefont {Zhu}}, \bibinfo
  {author} {\bibfnamefont {J.}~\bibnamefont {Szmelter}}, \ and\ \bibinfo
  {author} {\bibfnamefont {O.}~\bibnamefont {Zienkiewicz}},\ }\bibfield
  {title} {\enquote {\bibinfo {title} {Error estimation and adaptivity in
  navier-stokes incompressible flows},}\ }\href@noop {} {\bibfield  {journal}
  {\bibinfo  {journal} {Computational mechanics}\ }\textbf {\bibinfo {volume}
  {6}},\ \bibinfo {pages} {259--270} (\bibinfo {year} {1990})}\BibitemShut
  {NoStop}%
\bibitem [{\citenamefont {Leicht}\ and\ \citenamefont
  {Hartmann}(2010)}]{leicht2010error}%
  \BibitemOpen
  \bibfield  {author} {\bibinfo {author} {\bibfnamefont {T.}~\bibnamefont
  {Leicht}}\ and\ \bibinfo {author} {\bibfnamefont {R.}~\bibnamefont
  {Hartmann}},\ }\bibfield  {title} {\enquote {\bibinfo {title} {Error
  estimation and anisotropic mesh refinement for 3d laminar aerodynamic flow
  simulations},}\ }\href@noop {} {\bibfield  {journal} {\bibinfo  {journal}
  {Journal of Computational Physics}\ }\textbf {\bibinfo {volume} {229}},\
  \bibinfo {pages} {7344--7360} (\bibinfo {year} {2010})}\BibitemShut {NoStop}%
\bibitem [{\citenamefont {Pain}\ \emph {et~al.}(2001)\citenamefont {Pain},
  \citenamefont {Umpleby}, \citenamefont {De~Oliveira},\ and\ \citenamefont
  {Goddard}}]{pain2001tetrahedral}%
  \BibitemOpen
  \bibfield  {author} {\bibinfo {author} {\bibfnamefont {C.}~\bibnamefont
  {Pain}}, \bibinfo {author} {\bibfnamefont {A.}~\bibnamefont {Umpleby}},
  \bibinfo {author} {\bibfnamefont {C.}~\bibnamefont {De~Oliveira}}, \ and\
  \bibinfo {author} {\bibfnamefont {A.}~\bibnamefont {Goddard}},\ }\bibfield
  {title} {\enquote {\bibinfo {title} {Tetrahedral mesh optimisation and
  adaptivity for steady-state and transient finite element calculations},}\
  }\href@noop {} {\bibfield  {journal} {\bibinfo  {journal} {Computer Methods
  in Applied Mechanics and Engineering}\ }\textbf {\bibinfo {volume} {190}},\
  \bibinfo {pages} {3771--3796} (\bibinfo {year} {2001})}\BibitemShut {NoStop}%
\bibitem [{\citenamefont {Compere}, \citenamefont {Marchandise},\ and\
  \citenamefont {Remacle}(2008)}]{compere2008transient}%
  \BibitemOpen
  \bibfield  {author} {\bibinfo {author} {\bibfnamefont {G.}~\bibnamefont
  {Compere}}, \bibinfo {author} {\bibfnamefont {E.}~\bibnamefont
  {Marchandise}}, \ and\ \bibinfo {author} {\bibfnamefont {J.-F.}\ \bibnamefont
  {Remacle}},\ }\bibfield  {title} {\enquote {\bibinfo {title} {Transient
  adaptivity applied to two-phase incompressible flows},}\ }\href@noop {}
  {\bibfield  {journal} {\bibinfo  {journal} {Journal of Computational
  Physics}\ }\textbf {\bibinfo {volume} {227}},\ \bibinfo {pages} {1923--1942}
  (\bibinfo {year} {2008})}\BibitemShut {NoStop}%
\bibitem [{\citenamefont {Wissink}\ \emph {et~al.}(2010)\citenamefont
  {Wissink}, \citenamefont {Potsdam}, \citenamefont {Sankaran}, \citenamefont
  {Sitaraman}, \citenamefont {Yang},\ and\ \citenamefont
  {Mavriplis}}]{wissink2010coupled}%
  \BibitemOpen
  \bibfield  {author} {\bibinfo {author} {\bibfnamefont {A.}~\bibnamefont
  {Wissink}}, \bibinfo {author} {\bibfnamefont {M.}~\bibnamefont {Potsdam}},
  \bibinfo {author} {\bibfnamefont {V.}~\bibnamefont {Sankaran}}, \bibinfo
  {author} {\bibfnamefont {J.}~\bibnamefont {Sitaraman}}, \bibinfo {author}
  {\bibfnamefont {Z.}~\bibnamefont {Yang}}, \ and\ \bibinfo {author}
  {\bibfnamefont {D.}~\bibnamefont {Mavriplis}},\ }\bibfield  {title} {\enquote
  {\bibinfo {title} {A coupled unstructured-adaptive cartesian cfd approach for
  hover prediction},}\ }in\ \href@noop {} {\emph {\bibinfo {booktitle}
  {American Helicopter Society 66th Annual Forum}}}\ (\bibinfo {organization}
  {AHS International Alexandria, VA},\ \bibinfo {year} {2010})\ pp.\ \bibinfo
  {pages} {1300--1317}\BibitemShut {NoStop}%
\bibitem [{\citenamefont {Vidal}, \citenamefont {Wolf},\ and\ \citenamefont
  {Dupont}(2012)}]{vidal2012combinatorial}%
  \BibitemOpen
  \bibfield  {author} {\bibinfo {author} {\bibfnamefont {V.}~\bibnamefont
  {Vidal}}, \bibinfo {author} {\bibfnamefont {C.}~\bibnamefont {Wolf}}, \ and\
  \bibinfo {author} {\bibfnamefont {F.}~\bibnamefont {Dupont}},\ }\bibfield
  {title} {\enquote {\bibinfo {title} {Combinatorial mesh optimization},}\
  }\href@noop {} {\bibfield  {journal} {\bibinfo  {journal} {The Visual
  Computer}\ }\textbf {\bibinfo {volume} {28}},\ \bibinfo {pages} {511--525}
  (\bibinfo {year} {2012})}\BibitemShut {NoStop}%
\bibitem [{\citenamefont {Taira}\ and\ \citenamefont
  {Nair}(2022)}]{taira2022network}%
  \BibitemOpen
  \bibfield  {author} {\bibinfo {author} {\bibfnamefont {K.}~\bibnamefont
  {Taira}}\ and\ \bibinfo {author} {\bibfnamefont {A.~G.}\ \bibnamefont
  {Nair}},\ }\bibfield  {title} {\enquote {\bibinfo {title} {Network-based
  analysis of fluid flows: Progress and outlook},}\ }\href@noop {} {\bibfield
  {journal} {\bibinfo  {journal} {Progress in Aerospace Sciences}\ }\textbf
  {\bibinfo {volume} {131}},\ \bibinfo {pages} {100823} (\bibinfo {year}
  {2022})}\BibitemShut {NoStop}%
\bibitem [{\citenamefont {Taira}, \citenamefont {Nair},\ and\ \citenamefont
  {Brunton}(2016)}]{taira2016network}%
  \BibitemOpen
  \bibfield  {author} {\bibinfo {author} {\bibfnamefont {K.}~\bibnamefont
  {Taira}}, \bibinfo {author} {\bibfnamefont {A.~G.}\ \bibnamefont {Nair}}, \
  and\ \bibinfo {author} {\bibfnamefont {S.~L.}\ \bibnamefont {Brunton}},\
  }\bibfield  {title} {\enquote {\bibinfo {title} {Network structure of
  two-dimensional decaying isotropic turbulence},}\ }\href@noop {} {\bibfield
  {journal} {\bibinfo  {journal} {Journal of Fluid Mechanics}\ }\textbf
  {\bibinfo {volume} {795}},\ \bibinfo {pages} {R2} (\bibinfo {year}
  {2016})}\BibitemShut {NoStop}%
\bibitem [{\citenamefont {Meena}\ and\ \citenamefont
  {Taira}(2021)}]{meena2021identifying}%
  \BibitemOpen
  \bibfield  {author} {\bibinfo {author} {\bibfnamefont {M.~G.}\ \bibnamefont
  {Meena}}\ and\ \bibinfo {author} {\bibfnamefont {K.}~\bibnamefont {Taira}},\
  }\bibfield  {title} {\enquote {\bibinfo {title} {Identifying vortical network
  connectors for turbulent flow modification},}\ }\href@noop {} {\bibfield
  {journal} {\bibinfo  {journal} {Journal of Fluid Mechanics}\ }\textbf
  {\bibinfo {volume} {915}},\ \bibinfo {pages} {A10} (\bibinfo {year}
  {2021})}\BibitemShut {NoStop}%
\bibitem [{\citenamefont {Nair}\ and\ \citenamefont
  {Taira}(2015)}]{nair2015network}%
  \BibitemOpen
  \bibfield  {author} {\bibinfo {author} {\bibfnamefont {A.~G.}\ \bibnamefont
  {Nair}}\ and\ \bibinfo {author} {\bibfnamefont {K.}~\bibnamefont {Taira}},\
  }\bibfield  {title} {\enquote {\bibinfo {title} {Network-theoretic approach
  to sparsified discrete vortex dynamics},}\ }\href@noop {} {\bibfield
  {journal} {\bibinfo  {journal} {Journal of Fluid Mechanics}\ }\textbf
  {\bibinfo {volume} {768}},\ \bibinfo {pages} {549--571} (\bibinfo {year}
  {2015})}\BibitemShut {NoStop}%
\bibitem [{\citenamefont {Gopalakrishnan~Meena}, \citenamefont {Nair},\ and\
  \citenamefont {Taira}(2018)}]{gopalakrishnan2018network}%
  \BibitemOpen
  \bibfield  {author} {\bibinfo {author} {\bibfnamefont {M.}~\bibnamefont
  {Gopalakrishnan~Meena}}, \bibinfo {author} {\bibfnamefont {A.~G.}\
  \bibnamefont {Nair}}, \ and\ \bibinfo {author} {\bibfnamefont
  {K.}~\bibnamefont {Taira}},\ }\bibfield  {title} {\enquote {\bibinfo {title}
  {Network community-based model reduction for vortical flows},}\ }\href@noop
  {} {\bibfield  {journal} {\bibinfo  {journal} {Physical Review E}\ }\textbf
  {\bibinfo {volume} {97}},\ \bibinfo {pages} {063103} (\bibinfo {year}
  {2018})}\BibitemShut {NoStop}%
\bibitem [{\citenamefont {Nair}, \citenamefont {Douglass},\ and\ \citenamefont
  {Arya}(2023)}]{nair2023network}%
  \BibitemOpen
  \bibfield  {author} {\bibinfo {author} {\bibfnamefont {A.~G.}\ \bibnamefont
  {Nair}}, \bibinfo {author} {\bibfnamefont {S.~B.}\ \bibnamefont {Douglass}},
  \ and\ \bibinfo {author} {\bibfnamefont {N.}~\bibnamefont {Arya}},\
  }\bibfield  {title} {\enquote {\bibinfo {title} {Network-theoretic modeling
  of fluid--structure interactions},}\ }\href@noop {} {\bibfield  {journal}
  {\bibinfo  {journal} {Theoretical and Computational Fluid Dynamics}\ }\textbf
  {\bibinfo {volume} {37}},\ \bibinfo {pages} {707--723} (\bibinfo {year}
  {2023})}\BibitemShut {NoStop}%
\bibitem [{\citenamefont {Bai}\ \emph {et~al.}(2019)\citenamefont {Bai},
  \citenamefont {Erichson}, \citenamefont {Gopalakrishnan~Meena}, \citenamefont
  {Taira},\ and\ \citenamefont {Brunton}}]{bai2019randomized}%
  \BibitemOpen
  \bibfield  {author} {\bibinfo {author} {\bibfnamefont {Z.}~\bibnamefont
  {Bai}}, \bibinfo {author} {\bibfnamefont {N.~B.}\ \bibnamefont {Erichson}},
  \bibinfo {author} {\bibfnamefont {M.}~\bibnamefont {Gopalakrishnan~Meena}},
  \bibinfo {author} {\bibfnamefont {K.}~\bibnamefont {Taira}}, \ and\ \bibinfo
  {author} {\bibfnamefont {S.~L.}\ \bibnamefont {Brunton}},\ }\bibfield
  {title} {\enquote {\bibinfo {title} {Randomized methods to characterize
  large-scale vortical flow networks},}\ }\href@noop {} {\bibfield  {journal}
  {\bibinfo  {journal} {PloS one}\ }\textbf {\bibinfo {volume} {14}},\ \bibinfo
  {pages} {e0225265} (\bibinfo {year} {2019})}\BibitemShut {NoStop}%
\bibitem [{\citenamefont {Manohar}\ \emph {et~al.}(2018)\citenamefont
  {Manohar}, \citenamefont {Brunton}, \citenamefont {Kutz},\ and\ \citenamefont
  {Brunton}}]{manohar2018data}%
  \BibitemOpen
  \bibfield  {author} {\bibinfo {author} {\bibfnamefont {K.}~\bibnamefont
  {Manohar}}, \bibinfo {author} {\bibfnamefont {B.~W.}\ \bibnamefont
  {Brunton}}, \bibinfo {author} {\bibfnamefont {J.~N.}\ \bibnamefont {Kutz}}, \
  and\ \bibinfo {author} {\bibfnamefont {S.~L.}\ \bibnamefont {Brunton}},\
  }\bibfield  {title} {\enquote {\bibinfo {title} {Data-driven sparse sensor
  placement for reconstruction: Demonstrating the benefits of exploiting known
  patterns},}\ }\href@noop {} {\bibfield  {journal} {\bibinfo  {journal} {IEEE
  Control Systems Magazine}\ }\textbf {\bibinfo {volume} {38}},\ \bibinfo
  {pages} {63--86} (\bibinfo {year} {2018})}\BibitemShut {NoStop}%
\bibitem [{\citenamefont {Callaham}\ \emph {et~al.}(2021)\citenamefont
  {Callaham}, \citenamefont {Koch}, \citenamefont {Brunton}, \citenamefont
  {Kutz},\ and\ \citenamefont {Brunton}}]{callaham2021learning}%
  \BibitemOpen
  \bibfield  {author} {\bibinfo {author} {\bibfnamefont {J.~L.}\ \bibnamefont
  {Callaham}}, \bibinfo {author} {\bibfnamefont {J.~V.}\ \bibnamefont {Koch}},
  \bibinfo {author} {\bibfnamefont {B.~W.}\ \bibnamefont {Brunton}}, \bibinfo
  {author} {\bibfnamefont {J.~N.}\ \bibnamefont {Kutz}}, \ and\ \bibinfo
  {author} {\bibfnamefont {S.~L.}\ \bibnamefont {Brunton}},\ }\bibfield
  {title} {\enquote {\bibinfo {title} {Learning dominant physical processes
  with data-driven balance models},}\ }\href@noop {} {\bibfield  {journal}
  {\bibinfo  {journal} {Nature communications}\ }\textbf {\bibinfo {volume}
  {12}},\ \bibinfo {pages} {1016} (\bibinfo {year} {2021})}\BibitemShut
  {NoStop}%
\bibitem [{\citenamefont {Menon}\ and\ \citenamefont
  {Mittal}(2021)}]{menon2021significance}%
  \BibitemOpen
  \bibfield  {author} {\bibinfo {author} {\bibfnamefont {K.}~\bibnamefont
  {Menon}}\ and\ \bibinfo {author} {\bibfnamefont {R.}~\bibnamefont {Mittal}},\
  }\bibfield  {title} {\enquote {\bibinfo {title} {Significance of the
  strain-dominated region around a vortex on induced aerodynamic loads},}\
  }\href@noop {} {\bibfield  {journal} {\bibinfo  {journal} {Journal of Fluid
  Mechanics}\ }\textbf {\bibinfo {volume} {918}},\ \bibinfo {pages} {R3}
  (\bibinfo {year} {2021})}\BibitemShut {NoStop}%
\bibitem [{ofv(2023)}]{ofv2312}%
  \BibitemOpen
  \href@noop {} {\enquote {\bibinfo {title} {Openfoam version 2312},}\
  }\bibinfo {howpublished}
  {\url{https://www.openfoam.com/news/main-news/openfoam-v2312}} (\bibinfo
  {year} {2023})\BibitemShut {NoStop}%
\bibitem [{\citenamefont {Jakob}, \citenamefont {Rhinelander},\ and\
  \citenamefont {Moldovan}(2017)}]{pybind11}%
  \BibitemOpen
  \bibfield  {author} {\bibinfo {author} {\bibfnamefont {W.}~\bibnamefont
  {Jakob}}, \bibinfo {author} {\bibfnamefont {J.}~\bibnamefont {Rhinelander}},
  \ and\ \bibinfo {author} {\bibfnamefont {D.}~\bibnamefont {Moldovan}},\
  }\href@noop {} {\enquote {\bibinfo {title} {pybind11 -- seamless operability
  between c++11 and python},}\ } (\bibinfo {year} {2017}),\ \bibinfo {note}
  {https://github.com/pybind/pybind11}\BibitemShut {NoStop}%
\bibitem [{\citenamefont {Nicoud}\ and\ \citenamefont
  {Ducros}(1999)}]{nicoud1999subgrid}%
  \BibitemOpen
  \bibfield  {author} {\bibinfo {author} {\bibfnamefont {F.}~\bibnamefont
  {Nicoud}}\ and\ \bibinfo {author} {\bibfnamefont {F.}~\bibnamefont
  {Ducros}},\ }\bibfield  {title} {\enquote {\bibinfo {title} {Subgrid-scale
  stress modelling based on the square of the velocity gradient tensor},}\
  }\href@noop {} {\bibfield  {journal} {\bibinfo  {journal} {Flow, turbulence
  and Combustion}\ }\textbf {\bibinfo {volume} {62}},\ \bibinfo {pages}
  {183--200} (\bibinfo {year} {1999})}\BibitemShut {NoStop}%
\end{thebibliography}%
		
	\end{document}